\documentclass[lettersize,journal]{IEEEtran}
\usepackage{amsmath,amsfonts}
\usepackage{algorithmic}
\usepackage{algorithm}
\usepackage{array}
\usepackage[caption=false,font=normalsize,labelfont=sf,textfont=sf]{subfig}
\usepackage{textcomp}
\usepackage{stfloats}
\usepackage{url}
\usepackage{verbatim}
\usepackage{graphicx}
\usepackage{cite}
\usepackage{amssymb} 
\usepackage{mathrsfs}
\usepackage{mathtools}
\usepackage{bm}
\usepackage{cite}
\usepackage{color}

\newcommand{\Z}{\mathbb{Z}}
\newcommand{\R}{\mathbb{R}}

\newcommand{\Enc}{\mathsf{Enc}}

\newtheorem{remark}{Remark}

\begin{document}

\title{Encrypted Simultaneous Control of Joint Angle and Stiffness of Antagonistic Pneumatic Artificial Muscle Actuator by Polynomial Approximation}

\author{Yuta Takeda,
        Takaya Shin, 
        Kaoru Teranishi, and 
        Kiminao Kogiso

\thanks{This work was supported by JSPS KAKENHI Grant Numbers JP22H01509 and JP21K19762.}
\thanks{Y. Takeda, K. Teranishi, and K. Kogiso are with the Department of Mechanical and Intelligent Systems Engineering, 
The University of Electro-Communications,
1-5-1 Chofugaoka, Chofu, Tokyo 1828585, Japan.
e-mail: yutatakeda@uec.ac.jp.}%
\thanks{T. Shin is with DAIHEN, Inc.}}%

\maketitle

\begin{abstract}
This study proposes an encrypted simultaneous control system for an antagonistic pneumatic artificial muscle (PAM) actuator toward developing a cybersecure and flexible actuator.
First, a novel simultaneous control system design is considered for the joint angle and stiffness of a PAM actuator in a model-based design approach, facilitating the use of an encrypted control method.
The designed controller includes a contraction force model expressed as rational polynomial functions, which makes it difficult to encrypt the controller.
To overcome this difficulty, a least absolute shrinkage and selection operator (LASSO)-based polynomial approximation is employed for a rational controller.
The resulting polynomial controller is then transformed into a matrix-vector product form, which enables the use of a specific homomorphic encryption scheme to develop an encrypted simultaneous control system for the PAM actuator.
Finally, this study quantitatively evaluates the tracking control performance of the original, approximated, and encrypted controllers.
The experimental results show that the proposed encrypted controller achieves simultaneous tracking of the joint angle and stiffness with a tracking error of less than 2.7~\%.
\end{abstract}


\section{Introduction}
\IEEEPARstart{T}{he} McKibben pneumatic artificial muscle (PAM) was first developed in the 1950s \cite{1996_Chou}.
A PAM is inflated by injecting compressed air into a rubber tube wrapped in a nonstretchable mesh to generate a contraction force.
This structure offers several advantages over traditional motors and cylinders, including compact size, lightweight, and flexibility \cite{1995_Caldwell,2000_Tondu}.
Because PAMs produce contraction force in a single direction, an antagonistic configuration is typically adopted for practical applications, where two PAMs are arranged in parallel through joints \cite{10_Mihn, 2011_Beyl, 13_Andri, 2013_Tothova, 14_Andri, 2016_Andrikopoulos, 20_Jamwal, 2020_Gong, 2022_Zhou}. 
This structure facilitates rotational motion and closely mimics human movements \cite{2000_Tondu}.
Owing to these benefits, a review paper~\cite{2011_Andrikopoulos} stated that PAMs have found considerable use in the development of rehabilitation and assistive devices that can be comfortably worn and operated by users. 
Furthermore, numerous studies have explored stiffness or compliance control \cite{2010_stiffness, 2012_Sui, 2015_Ugurlu, 2017_okajima, 2017_Zhao} and developed methods of simultaneous control of the joint angle and stiffness to leverage the potential of PAMs fully \cite{2009_nakamura, 2011_Choi, 2018_Cao, 2016_saito, 2019_Ugurl, 2022_shin_reference}.
These active controls enhance safety during contact and collision between robots and humans or their environments while also augmenting the comfort of wearable devices.

Applications that integrate wireless network technology into PAM actuator systems include remote surgical robots \cite{2012_Li}, remotely operated cranes \cite{2006_Sasaki}, teleoperation of pneumatic robots \cite{2010_Kato}, and walking-assistive devices \cite{remote_PAM,Yean_20}.
The use of wireless communications enhances the portability of devices and promotes the self-rehabilitation of patients at home.
In self-rehabilitation, the monitoring of wearable devices through a network by physical therapists helps prevent injury due to falls.
Meanwhile, attention must be paid to the cybersecurity of networked actuator systems.
Cyberattack incidents and risk analyses have been reported, such as the falsification of control parameters by Stuxnet to destroy process plants \cite{2011_Langner},
an unmanned aerial vehicle compromised by hijacking video streaming \cite{2009_Gorman},
and the compromising of a robot controller \cite{2017_Quarta}.

To develop secure PAM actuator systems, networked actuator systems must be equipped with cybersecurity countermeasures.
Encrypted control was proposed as a cyberserucrity countermeasure for networked control systems in \cite{2015_Kogiso} by integrating the homomorphism of a specific public-key encryption scheme into a linear or polynomial-type control system.
Encrypted control conceals the control parameters inside the control device and the signals over the communication links, resulting in protection against eavesdropping \cite{2015_Kogiso, Ter20AIM, 2019_Qiu, 2022_shin_cyber} and real-time detection of cyberattacks \cite{Baba18_2,2015_Kogiso,MiyamotoAccess}.
Encrypted implementation is expected to enhance the cybersecurity of control systems.

However, designing a nonlinear controller that is not a polynomial, such as a rational polynomial or switching controllers, is difficult.
A simultaneous control system for the joint angle and stiffness of an antagonistic PAM actuator uses a rational polynomial-type controller \cite{2009_nakamura, 2016_saito, 2019_Ugurl}. 
In additional, encrypted proportional-integral (PI) control of the PAM joint angle or torque was considered in \cite{2022_shin_cyber}, and in \cite{2020_Darup}, the implementation of encrypted polynomial controllers was demonstrated, verifying its effectiveness through a numerical toy example.
Unfortunately, to the best of our knowledge, no studies have been conducted on encrypting such nonlinear controllers.

The objective of this study is to propose an encrypted simultaneous control system for an antagonistic PAM actuator to develop a cybersecure and flexible actuator.
In this study, the simultaneous tracking control of the joint angle and stiffness of the antagonistic PAM actuator is considered for a step-like reference.
We present a novel model-based nonlinear control system comprising three components: a reference generator, a contraction force estimator, and PI controllers.
However, a model-based controller includes specific rational functions, and applying the encrypted control method is challenging.
To overcome this difficulty, this study approximates the designed controller as a polynomial-type controller that is friendly to controller encryption.
We then quantitatively investigate the effects of polynomial approximation and controller encryption on the control performance to evaluate the developed encrypted control system.
Finally, an experimental investigation confirms that the proposed encrypted control system enables simultaneous tracking to the step-like reference within an acceptable tracking error.

This study makes the following main contributions: 
it is the first to present an encrypted nonlinear control system for simultaneously tracking the joint angle and stiffness of an antagonistic PAM actuator, and
it provides a novel form of the contraction force model, which helps reduce the number of approximations that cause control performance degradation throughout the controller encryption procedure.

The remainder of this paper is organized as follows. 
Section~\textbf{\ref{sec:simucon}} introduces the antagonistic PAM actuator with a control objective and presents a model-based simultaneous control method for tracking the joint angle and stiffness of the PAM actuator to the reference.
Section~\textbf{\ref{sec:polyapprox}} discusses the polynomial approximation to acquire a friendly form of the controller to realize the encrypted control.
Section~\textbf{\ref{sec:ec}} proposes an encrypted-controlled PAM actuator to maintain the controller and communication secrets and investigates the impact of the secure implementation on the control performance to highlight the effectiveness of the proposed encrypted controller.
Finally, Section~\textbf{\ref{sec:con}} concludes the study.

\begin{figure}[t]
	\centering
 	\subfloat[]{\includegraphics[width=0.4\hsize]{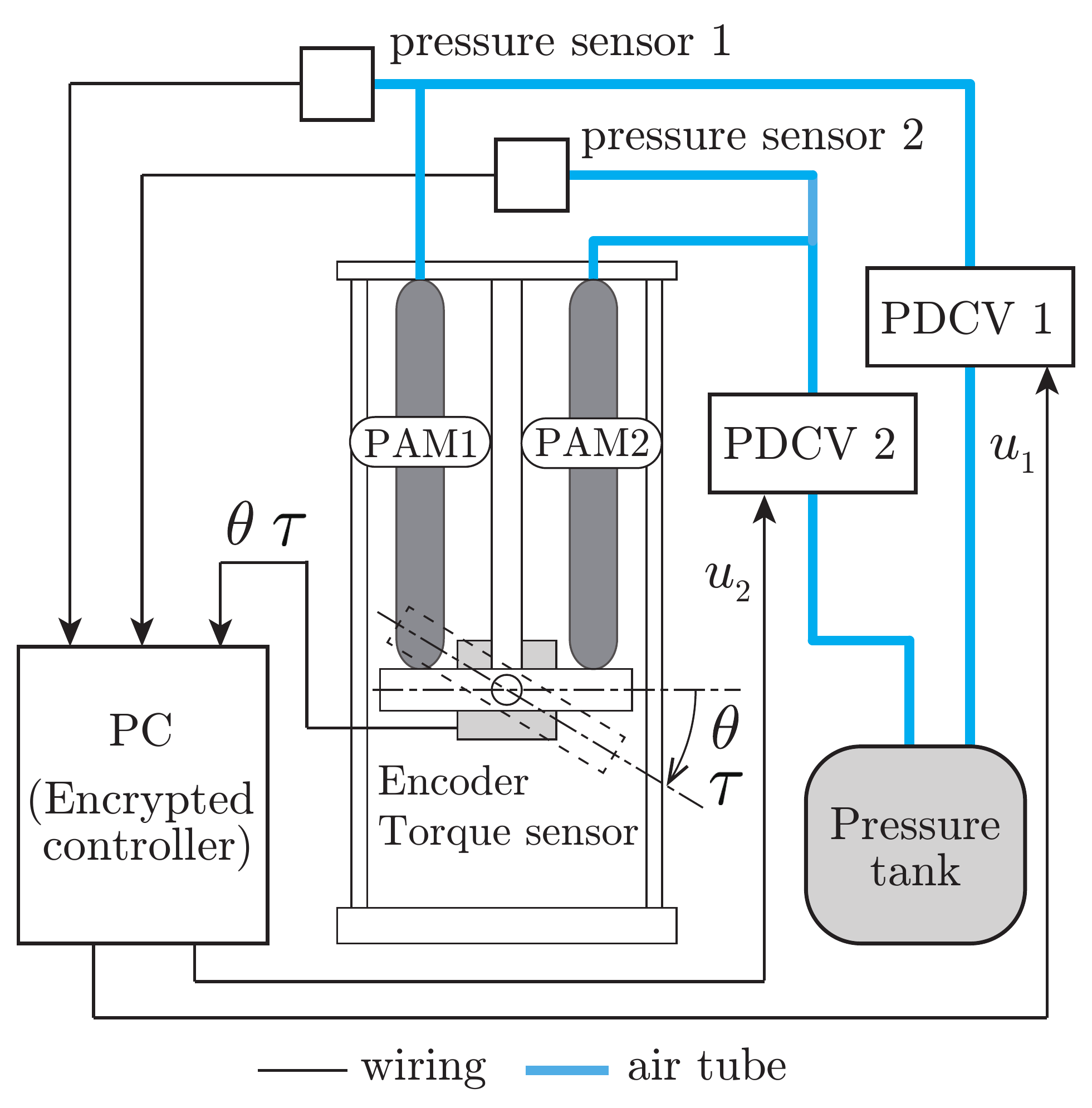}%
	\label{sfig:eqatg}}\qquad
    \subfloat[]{\includegraphics[scale=0.28]{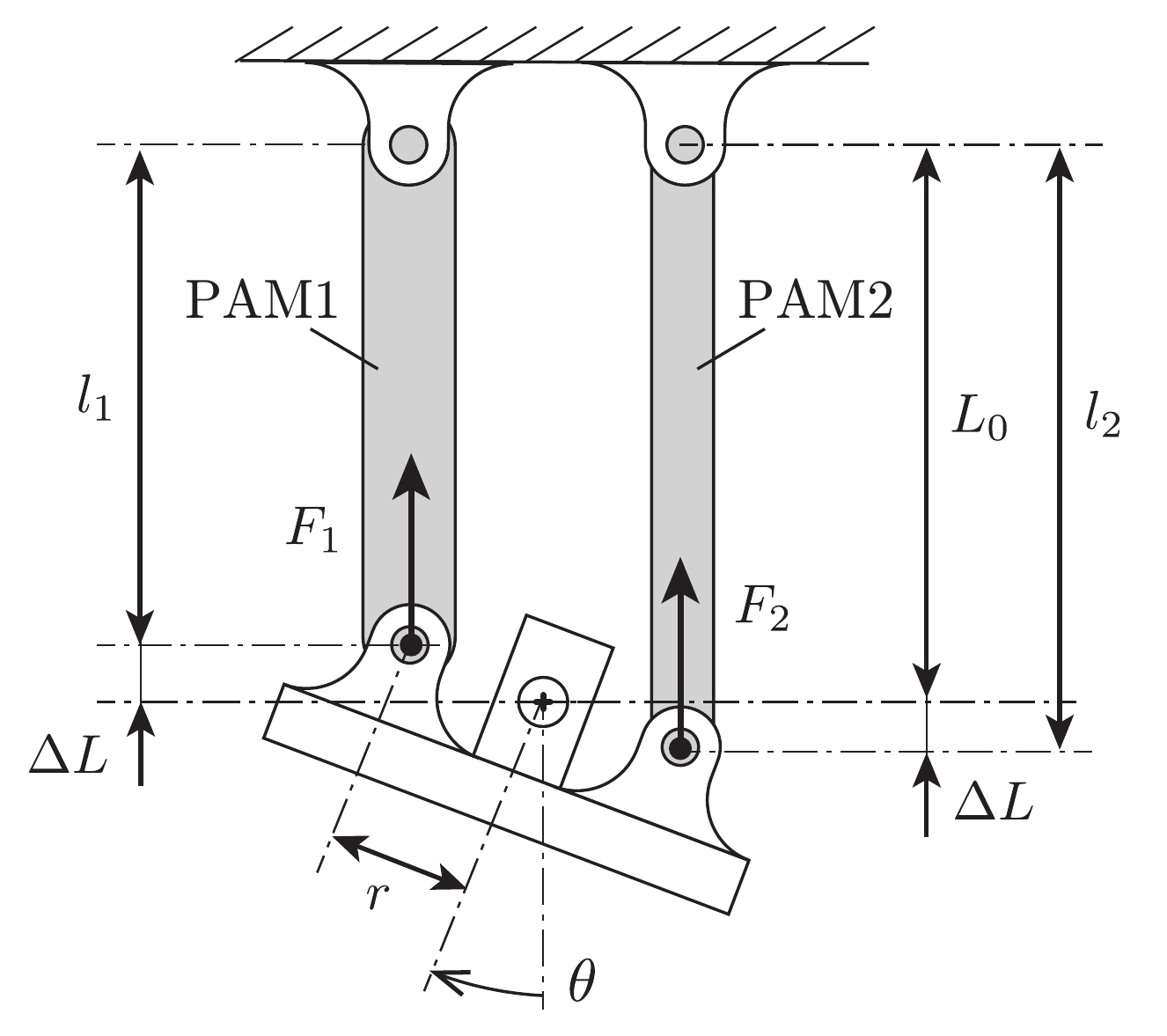}\label{fig:apam}}
	\caption{Antagonistic PAM actuator system considered \cite{2022_shin_model}. (a) Schematic of the entire system and (b) geometric structure of the PAM joint~\cite{2022_shin_reference}.}
	\label{fig:setup}
\end{figure}

\begin{table}[t]
		\centering
		\caption{Experimental equipments}
		\renewcommand{\arraystretch}{1}
		\begin{tabular}{ll}
		\hline
		\hline
		Name & Specifications\\
		\hline
		PAM & AirMuscle,  Kanda Tsushin Kogyo,\\
			& Length: 170 mm (be given a 2 kgf mass),\\
			& Diameter: 0.5 inches\\
		Valve (PDCV) & 5/3-way valve, FESTO,\\
			& a critical frequency of 125 Hz. \\
		Torque sensor& UTM II-10  Nm(R), UNIPULSE,\\
			&  range: $\pm 10$ Nm. \\
		Encoder& UTM II-10  Nm(R), UNIPULSE,\\
			&  Incremental, resolution: 2000 P/R\\
		Pressure sensor & E8F2-B10C, OMRON,\\
			& Range: $0$--$1$ MPa. \\
		Air compressor & 6-25, JUN-AIR, \\
			& Tank: 25 L; displacement: 60 L/min. \\
		Pressure tank & AST-25G, EARTH MAN, 25 L. \\
		PC & Ubuntu12.04, Xenomai2.6.2.1 Patch, \\
			& CPU: 3.2 GHz, memory: 8 GB. \\
		\hline
		\hline
		\end{tabular}
		\label{tbl:atgeq}
\end{table}

\section{Nonlinear Control System Design for Antagonistic PAM Actuator}\label{sec:simucon}
This section introduces the antagonistic PAM actuator, its simultaneous control system desingn, and the numerical evaluation of the control performance of the designed control system.

\subsection{Antagonistic PAM Actuator System}
Fig.~\ref{fig:setup}\subref{sfig:eqatg} shows an antagonistic PAM system.
An antagonistic PAM actuator is a joint actuator driven by two PAMs.
One side of each PAM is connected by a joint.
The system consists of two PAMs, two proportional directional control valves (PDCVs), an air tank, pressure sensors, a torque meter, a rotary encoder, and a control PC.
The tank stores compressed air connected to the PDCVs and PAMs using air tubes.
The airflow regulated by the PDCVs drives the PAMs, and the joint rotates.
The system inputs are the voltage commands to the two PDCVs ($u_1$ and $u_2$), and the measured values are the joint angle $\theta$, torque $\tau$, and inner pressure of the PAMs ($P_1$ and $P_2$).
The range of the rotation angle is $\pm$ 25~deg, and the range of the output torque is $\pm$ 3.0~Nm.
The sampling period $T_{\rm s}$ was set to 20~ms. 
Table \ref{tbl:atgeq} lists the details of the experimental equipment used.

The control objective considered in this study is to track the joint angle and stiffness of the PAM actuator simultaneously with respect to a given reference.
The goal of this study is to develop an encrypted simultaneous tracking control system for the joint angle and stiffness of a PAM actuator to mitigate the control performance degradation caused by secure implementation.

\subsection{Joint Stiffness}\label{ssec:stiff}
Joint stiffness is an index of the difficulty of a joint rotating against external forces \cite{2019_Ugurl,2022_shin_reference}.
This study employs the model of joint stiffness presented in \cite{2022_shin_reference}, following the introduction of the notations of the physical variables regarding the PAM.
The geometric relationships of the antagonistic PAM actuator are illustrated in Fig.~\ref{fig:setup}\subref{fig:apam}, the lengths of the two PAMs, $l_1$ and $l_2$, are respectively given by
$l_1(k)=L_0-\Delta L(k)$ and $l_2(k)=L_0+\Delta L(k)$, where $k\in\mathbb{Z}^+:=\{1,2,\cdots\}$ is a step, $\Delta L (k)\approx r\sin\theta(k)$ is the horizontal displacement of the two PAMs, $r$ is the radius of the joint, and $L_0$ is the PAM length when the joint is at the horizontal position. 
The contraction force of the PAM $F_i$, $\forall i\in{\mathcal I}:=\{1,2\}$ can be expressed as a function of the inner pressure and length as follows:
\begin{align}
	F_i(l_i(k),P_i(k))=a_i(l_i(k))P_i(k)+b_i(l_i(k)),\ \,\forall i\in\mathcal{I},
	\label{eq:force}
\end{align}
with $a_i(l)=p^{a_1}_il+p^{a_2}_i$ and $b_i(l)=p^{b_1}_il+p^{b_2}_i,\,\forall i\in\mathcal{I}$. 

When the joint torque $\tau$ generated by the two PAMs is given by $\tau(k) = r\cos \theta(k) \left( F_1(k) - F_2(k) \right)$, 
the joint stiffness $K_{P}$ is defined as the partial differentiation of joint torque $\tau$ by joint angle $\theta$ as follows \cite{2022_shin_reference}:
\begin{align}
 K_{P}(k) &=r\sin\theta(k) (F_1(k) - F_2(k)) \label{eq:K} \\
 &\quad +r^2\cos^2\theta(k) \left(\frac{F_1(k)-\alpha_1(k)}{l_1(k)}+\frac{F_2(k)-\alpha_2(k)}{l_2(k)}\right),\notag
\end{align}
where $\alpha_i(k) := p^{a_2}_iP_i(k)+p^{b_2}_i,\,\forall i\in\mathcal{I}$. 

\begin{figure}[tb]
	\centering
	\includegraphics[width=1\hsize]{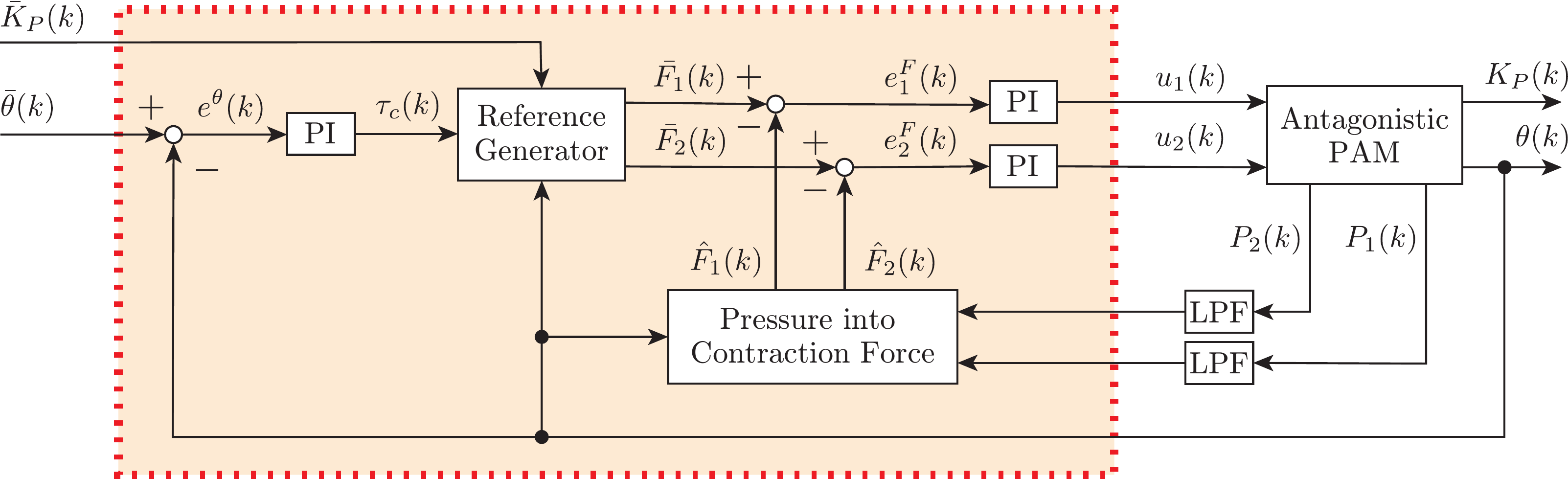}
	\caption{Block diagram of angle-stiffness control system.}
	\label{fig:stiff_controller}
\end{figure}

\subsection{Angle-Stiffness Controller}\label{sec:cont}
This study proposes a novel nonlinear controller for the simultaneous control of joint angle and stiffness, which is highlighted in red in Fig.~\ref{fig:stiff_controller}, as follows:
\begin{subequations}
\label{eq:ctrller}
\begin{align}
x(k+1)&=A(k;\theta)x(k)+g(k;\zeta),
\label{eq:controllera}\\
\begin{bmatrix}
u_1(k) \\ u_2(k)
\end{bmatrix}&=C(k;\theta)x(k)+h(k;\zeta)+
\begin{bmatrix}
    \beta_1 \\ \beta_2
\end{bmatrix},\label{eq:controllerb}
\end{align}
\end{subequations}
where $x\in\R^3$ is the state, 
$u_1$ and $u_2$ are the outputs, and $\zeta\in\R^5$ is the input, denoted by $\zeta:=[\,P_1\ \, P_2\ \,\theta\ \,\bar\theta\ \,\bar{K}_P\,]^\mathrm{T}$, where $P_1$, $P_2$, and $\theta$ can be measured by the sensor, and $\bar{\theta}$ and $\bar{K}_P$ are the references.
$\beta_1$ and $\beta_2$ are the bias voltages of each PDCV.
$A(\theta)\in\R^{3\times 3}$ and $C(\theta)\in\R^{2\times 3}$ are time-varying, $\theta$-dependent coefficients, and $g:\R^5\rightarrow\R^3$ and $f:\R^5\rightarrow\R^3$ are nonlinear functions of $\zeta$.
The coefficients and nonlinear functions are clarified after introducing a reference generator, pressure into the contraction force, and PI controllers, which are components of the proposed controller and are explained as follows.

\subsubsection{Reference Generator}
The reference generator outputs the reference signals of the contraction force for two PAMs, denoted as $\bar{F}_1$ and $\bar{F}_2$, taking the measured joint angle $\theta$, computed torque command $\tau_c$, and the given reference of joint stiffness $\bar{K}_P$ as inputs.
Using \cite{2022_shin_reference}, $\bar{F}_1$ and $\bar{F}_2$ are expressed as rational polynomial functions of the inputs $\theta$, $\tau_c$, and $\bar{K}_P$:
\begin{subequations}
\begin{align}
	\bar{F}_{1}(k) = &\ \frac{1}{r^2\cos^2\theta(k)}\frac{l_1(k)l_2(k)}{l_1(k) + l_2(k)}\biggl[ \bar{K}_P(k) \nonumber \biggr.
	+ \Bigl(\frac{r\cos\theta(k)}{l_2(k)}-\\ & \tan\theta(k)\Bigr)\tau_c(k)
	 \biggl.- r^2\cos^2\theta(k)\Bigl(\frac{\alpha_1(k)}{l_1(k)} + \frac{\alpha_2(k)}{l_2(k)}\Bigr) \biggr], \label{eq:F1_generate}\\
	\bar{F}_{2}(k) = &\ \bar{F}_{1}(k) - \frac{\tau_c(k)}{r\cos\theta(k)}.
	\label{eq:F2_generate}
\end{align}
\label{eq:barF}
\end{subequations}

\subsubsection{Pressure into Contraction Force}
The pressure-to-contraction force block shown in Fig.~\ref{fig:stiff_controller} estimates the contraction force of the PAM by using the following experimental polynomial function of the inner pressure and measured joint angle:
\begin{subequations}
\begin{align}
	\hat{F}_i&(\theta(k),P_i(k))=\hat{a}_i(\theta(k)) P_i(k)+\hat{b}_i(\theta(k)),\label{eq:esti_F1}\\
 &\hat{a}_i(\theta)=p^{\hat{a}_1}_i\theta+p^{\hat{a}_2}_i, \quad 
\hat{b}_i(\theta)=p^{\hat{b}_1}_i\theta+p^{\hat{b}_2}_i,
 \label{eq:esti_F2}
\end{align}
\label{eq:esti_F}
\end{subequations}
\!\!where $\hat{F}_i$, $\hat{a}_i$, and $\hat{b}_i$, $\forall i\in\mathcal{I}$, are an estimate of the PAM contraction force and the coefficients of \eqref{eq:esti_F1}, respectively.

\begin{figure}[tbp]
\centering
	\subfloat[]{\includegraphics[scale=.33]{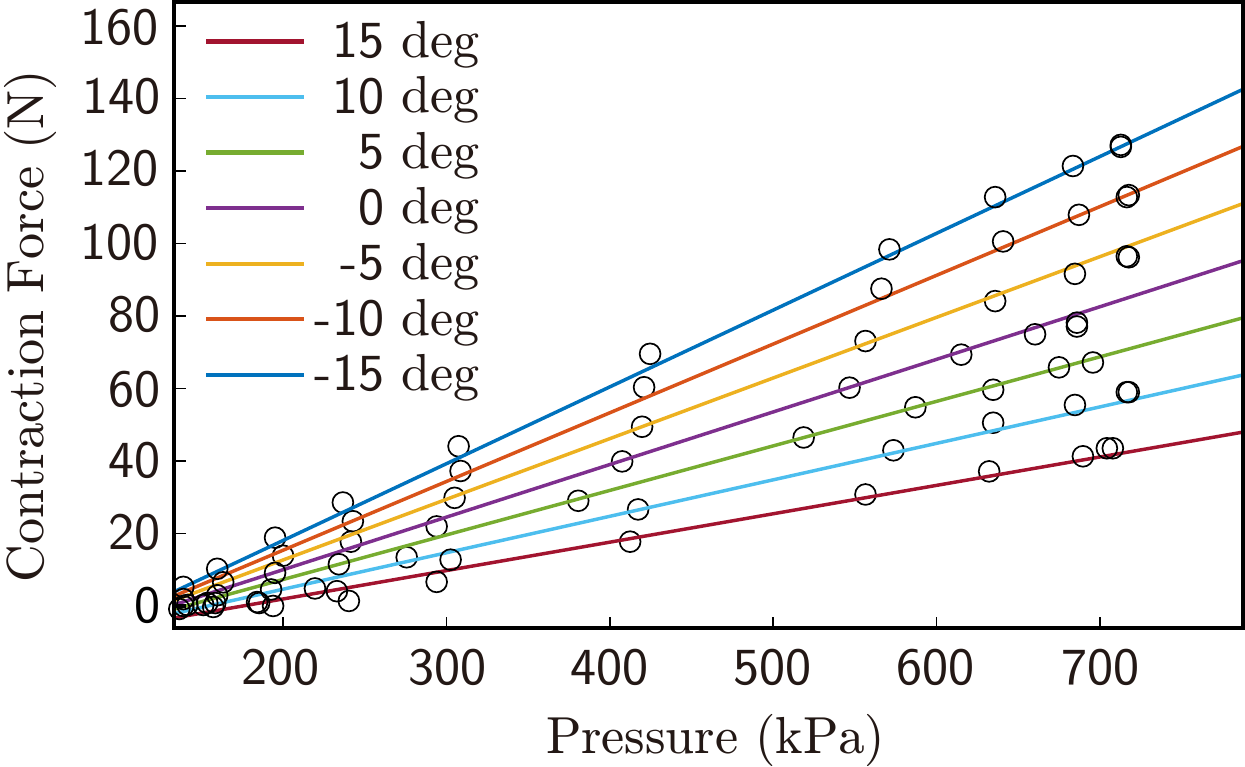}\label{sfig:P-F}}\ \ 
 	\subfloat[]{\includegraphics[scale=.33]{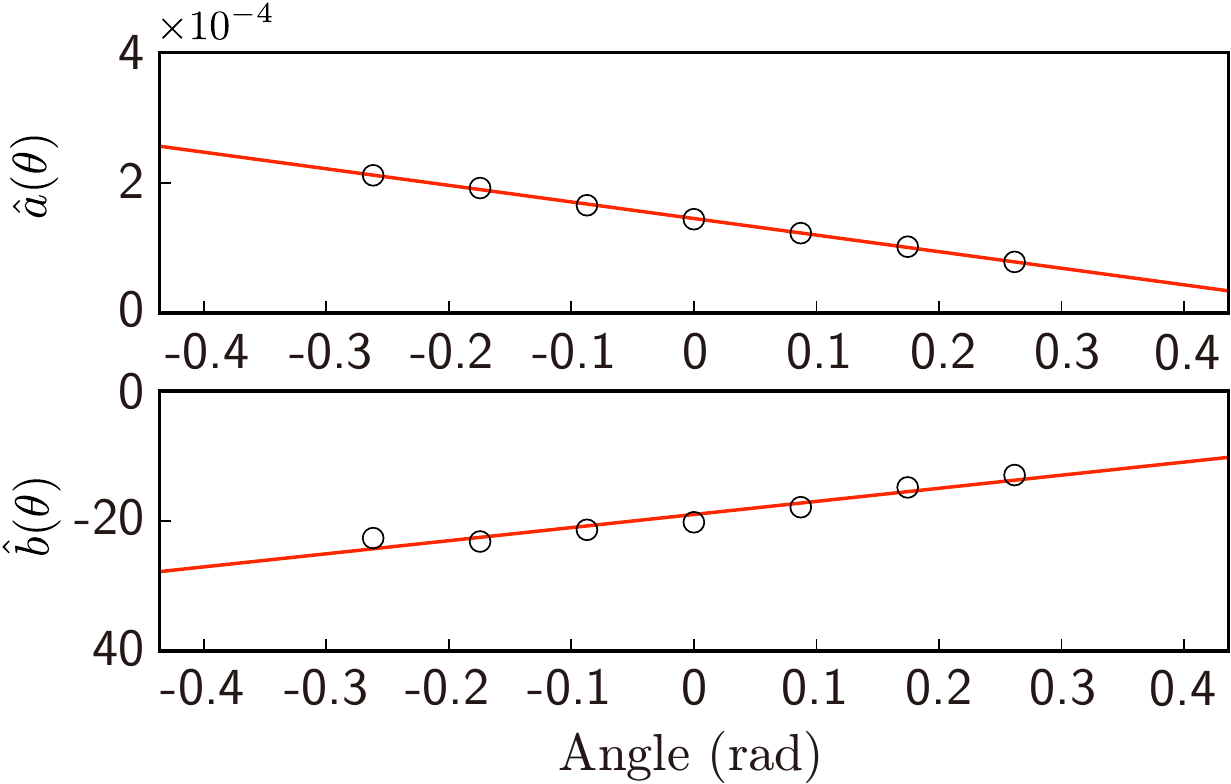}\label{sfig:v-w}}
    \caption{Fitting result of the $\theta$-dependent model \eqref{eq:esti_F} for capturing static characteristics with regard to $P$, $F$, and $\theta$ of the PAMs. (a) $P$-$F$ relationship of the PAM. (b) $\hat{a}$-$\theta$ and $\hat{b}$-$\theta$ relationships of the PAM.}
	\label{fig:new_force}
\end{figure}

A reason for introducing the contraction force model \eqref{eq:esti_F} in addition to \eqref{eq:force} is to reduce the number of approximation, which will facilitate the discussion in Section \ref{sec:polyapprox}.
Model \eqref{eq:force} involves a nonlinear term $\sin\theta$, which requires the approximation process to be a polynomial. 
On the other hand, the model \eqref{eq:esti_F} is a polynomial of $\theta$ and captures a static relationship between the angle and the contraction force.
Indeed, the fitting results of \eqref{eq:esti_F} are shown in Fig.~\ref{fig:new_force}.
In Fig.~\ref{fig:new_force}\subref{sfig:P-F}, the black circles represent experimentally measured data, and the colored lines represent fitting results using \eqref{eq:esti_F} for each angle. 
In this case, the coefficients \eqref{eq:esti_F2} for each angle correspond to Fig.~\ref{fig:new_force}\subref{sfig:v-w}, where the black circles represent the coefficients corresponding to the colored lines in Fig.~\ref{fig:new_force}\subref{sfig:P-F}.
The fitting test confirms that the $\theta$-dependent model \eqref{eq:esti_F} is valid.

\subsubsection{PI Controller}
The proposed controller includes two types of PI controllers to compensate for errors in the joint angle and the contraction force.
The controller for the joint angle uses a feedback error between $\bar{\theta}$ and $\theta$, denoted by $e^\theta:=\bar{\theta}-\theta$, to compute the command torque $\tau_{c}$, as follows:
\begin{align}
	\begin{bmatrix}
		x^\theta(k+1)\\
		 \tau_c(k)
	\end{bmatrix}=
	\begin{bmatrix}
		 1 & T_{s}\\
	    	G_{\rm I}^\theta & G_{\rm P}^\theta
	\end{bmatrix}
	\begin{bmatrix}
		x^\theta(k)\\
		e^\theta(k)
	\end{bmatrix},
\label{eq:PI1}
\end{align}
where $x^\theta$ is the state and $G_{\rm P}^\theta$ and $G_{\rm I}^\theta$ are the proportional and integral gains, respectively.

Regarding the contraction force controller, 
the outputs of the reference generator, $\bar{F}_1$ and $\bar{F}_2$, are compared with the estimated forces $\hat{F_1}$ and $\hat{F_2}$ computed by \eqref{eq:esti_F}, respectively.
The resulting errors are fed to the PI controllers to generate the control input voltages $u_1$ and $u_2$ for each PDCV.
The controllers are given as follows:
\begin{align}
	\begin{bmatrix}
		x_i^F(k+1)\\
		u_i(k)
	\end{bmatrix}=
	\begin{bmatrix}
		 1 & T_{s}&0\\
	    	G_{\rm I}^F & G_{\rm P}^F&1
	\end{bmatrix}
	\begin{bmatrix}
		x_i^F(k)\\
		e_i^F(k)\\
		\beta_i
	\end{bmatrix},\ \forall i\in\mathcal{I},
\label{eq:PI2}
\end{align}
where $x_i^F, \forall i\in\mathcal{I},$ is the state of the $i$-th controller, $e_i^F:= \bar{F}_i-\hat{F}_i, \forall i\in\mathcal{I},$ is the error in the contraction force of the $i$th PAM, $G_{\rm P}^F$ and $G_{\rm I}^F$ are the proportional and integral gains, respectively, and the gains are common between the controllers.
Additionally, we set $\beta_i=5.0, \forall i\in\mathcal{I}$.

Consequently, defining the controller state by $x:=[\,x^\theta\ \,x_1^F\ \,x_2^F\,]^\mathrm{T}$ and eliminating $\tau_c$, $e^\theta$, $e^F_i$, $\bar{F}_i$, and $\hat{F}_i$, $\forall i\in\mathcal{I}$ from \eqref{eq:barF}--\eqref{eq:PI2} results in the nonlinear controller \eqref{eq:ctrller} with the coefficients and functions in \eqref{eq:ACgh}. 

\begin{figure*}[t]
\begin{subequations}
\label{eq:ACgh}
\begin{align}
A(k;\theta)=&
\begin{bmatrix}
1 & 0 & 0\\
\frac{T_sG_\textrm{I}^\theta l_1(k)l_2(k)}{r^2(l_1(k)+l_2(k))\cos^2\theta(k)}\Bigl(\frac{r\cos\theta(k)}{l_2(k)}-\tan\theta(k)\Bigr) & 1 & 0\\
T_sG_\textrm{I}^\theta\left\{\frac{l_1(k)l_2(k)}{r^2(l_1(k)+l_2(k))\cos^2\theta(k)}\Bigl(\frac{r\cos\theta(k)}{l_2(k)}-\tan\theta(k)\Bigr)-\frac{1}{r\cos\theta(k)}\right\} & 0 & 1
\end{bmatrix},\quad
g(k;\zeta)=T_s
\begin{bmatrix}
\bar{\theta}(k)-\theta(k)\\
h_1(k;\zeta) \\ h_2(k;\zeta)
\end{bmatrix}, \label{ACgh_A}\\[1ex]
C(k;\theta)=&
\begin{bmatrix}
G_\textrm{P}^FG_\textrm{I}^\theta & G_\textrm{I}^F & 0 \\
G_\textrm{P}^FG_\textrm{I}^\theta\left\{\frac{l_1(k)l_2(k)}{r^2(l_1(k)+l_2(k)) \cos^2\theta(k)}\Bigl(\frac{r\cos\theta(k)}{l_2(k)}-\tan\theta(k)\Bigr)-\frac{1}{r\cos\theta(k)}\right\} & 0 & G_I^F
\end{bmatrix},\quad
h(k;\zeta)=G_\textrm{P}^F
\begin{bmatrix}
    h_1(k;\zeta) \\ h_2(k;\zeta)
\end{bmatrix},\label{ACgh_C}\\[1ex]
h_1(k;\zeta):=&\frac{l_1(k)l_2(k)}{l_1(k)+l_2(k)}\left[\frac{1}{r^2\cos^2\theta(k)}\left\{-\bar{K}_P(k)+G_\textrm{P}^\theta\Bigl(\frac{r\cos\theta(t)}{l_2(k)}-\tan\theta(k)\Bigr)(\bar{\theta}(k)-\theta(k))\right\}+\frac{\alpha_1(k)}{l_1(k)}+\frac{\alpha_2(k)}{l_2(k)}\right]-\notag\\
&\hat{a}_1P_1(k)-\hat{b}_1,\label{eq:ACgh_h1}\\[1ex]
h_2(k;\zeta):=&\frac{l_1(k)l_2(k)}{l_1(k)+l_2(k)}\left[\frac{1}{r^2\cos^2\theta(k)}\left\{-\bar{K}_P(k)+G_\textrm{P}^\theta\Bigl(\frac{r\cos\theta(k)}{l_2(k)}-\tan\theta(k)-\frac{l_1(k)+l_2(k)}{l_1(k)l_2(k)}r\cos\theta(k)\Bigr)(\bar{\theta}(k)-\theta(k))\right\}+\right.\notag\\
&\left.\frac{\alpha_1(k)}{l_1(k)}+\frac{\alpha_2(k)}{l_2(k)}\right]-\hat{a}_2P_2(k)-\hat{b}_2,\label{eq:ACgh_h2}
\end{align}
\end{subequations}
\hrulefill
\end{figure*}

\begin{figure*}[htbp]
\centering
	\subfloat[]{\includegraphics[scale=.53]{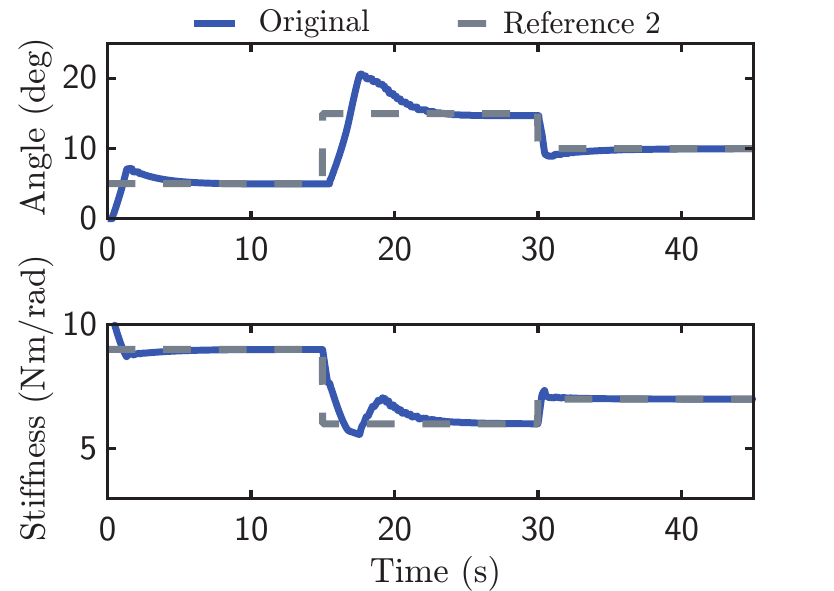}\label{sfig:ori2_out}}
 	\subfloat[]{\includegraphics[scale=.53]{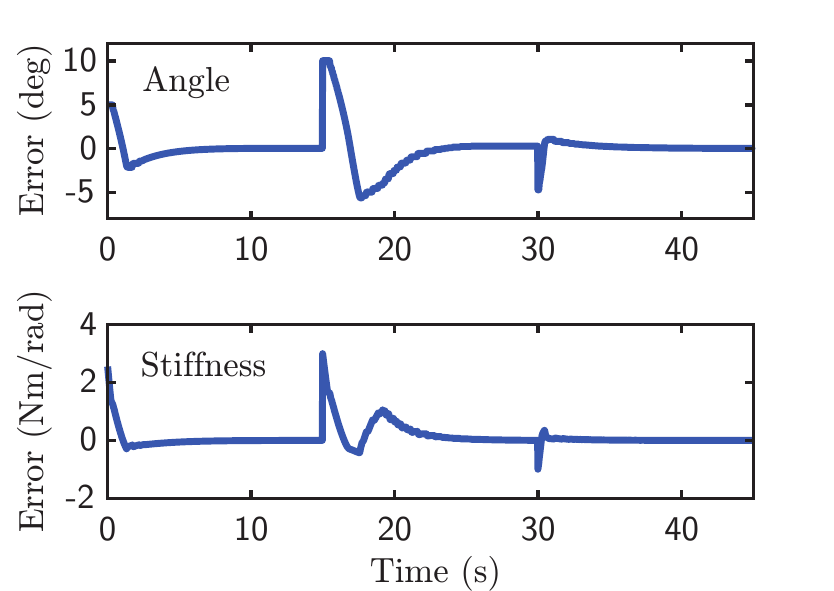}\label{sfig:ori2_error}}
 	\subfloat[]{\includegraphics[scale=.53]{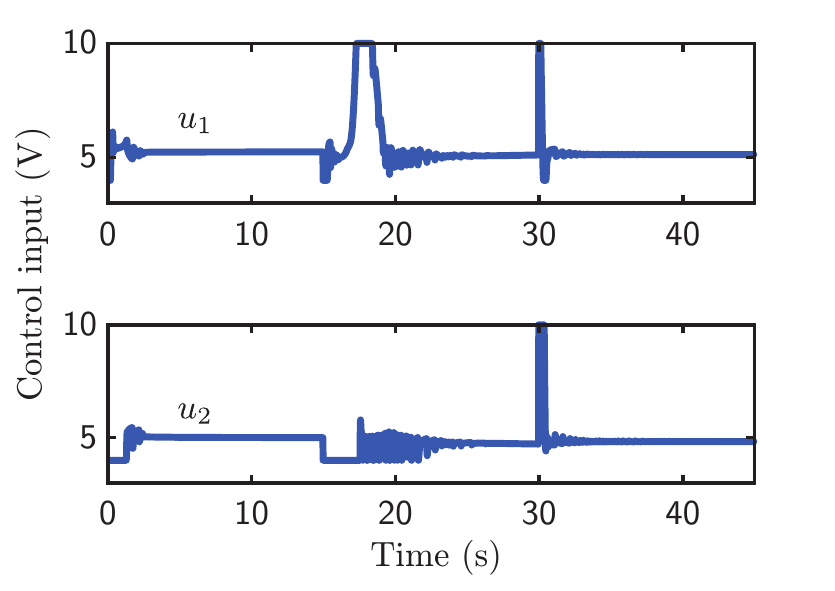}\label{sfig:ori2_ele}}
  	\subfloat[]{\includegraphics[scale=.53]{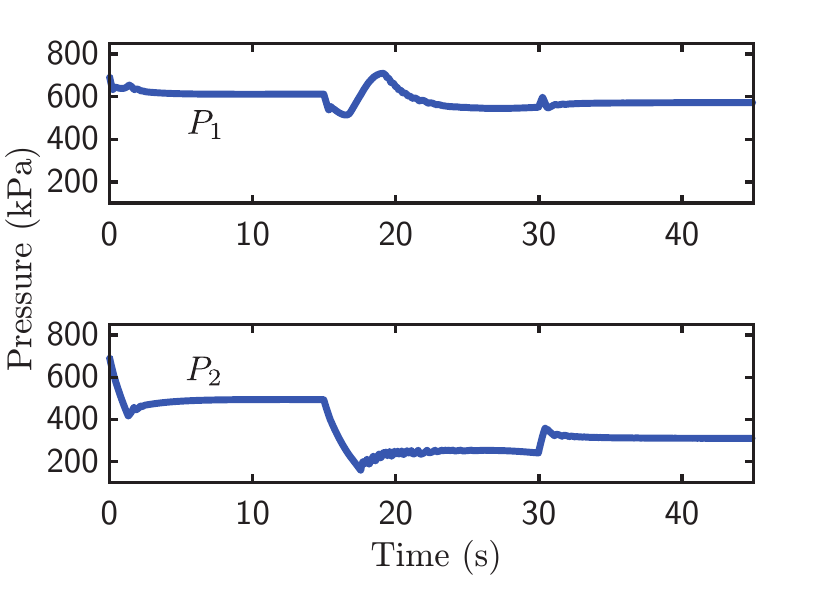}\label{sfig:ori2_pre}}
    \caption{Simulation results of the original simultaneous control of joint angle and stiffness for Reference~2. (a) Joint angle and stiffness, (b) tracking errors of the joint angle and stiffness, (c) control inputs to the valve, and (d) inner pressure of each PAM.}\vspace{-2ex}
    \label{fig:sim2}
\end{figure*}

\subsection{Numerical Verification}
This section presents the numerical simulations conducted to evaluate the proposed nonlinear controller in terms of the control performance with respect to simultaneous angle-stiffness control.
For the simulations, we used the following state-space model of an antagonistic PAM system \cite{2022_shin_model}: 
$x_p(k+1)=f_\sigma(x_p(k),u(k))$ if $x_p(k)\in\mathcal{X}_{\sigma}$ holds, 
$y(k)=h_p(x_p(k))$, 
where $u:=[u_1 \ u_2]^\mathrm{T}\in[0,10]^2\subset\R^2$ is the input voltages (V) to the two PDCVs, 
the state variable is $x_p:=[\,\theta\,\ \dot{\theta}\,\ P_1\,\ P_2 \,]^\mathrm{T}\in\R^4$, and the output variable is $y:=[\,\theta\,\ P_1\,\ P_2\,\ K_P\,]^\mathrm{T}\in\R^4$, in which a set of allowable absolute pressures (kPa) is determined by the specification of the PAMs, i.e., $P_1, P_2\in[200,750]$, 
$f_{\sigma}:\R^4\rightarrow\R^4$ is a nonlinear function with 18 subsystems, and it switches according to if-then rules,
${\mathcal X}_\sigma:=\{x_p\in\R^4\,|\,\Psi_{\sigma}(x_p)>0\}$ is the state set, where $\sigma\in\{1,2,\cdots,18\}$ is the index of the subsystem, 
$\Psi_{\sigma}(x_p)$ is a function derived from the modes in the form of if-then rules,
and the function $h_p:\R^4\rightarrow\R^4$ is an observation equation.

We considered two types of references of the joint angle and stiffness. 
Reference~1 was set to
\begin{align*}
(\bar{\theta}(k),\bar{K}_P(k))=&\ 
\begin{cases}
\,(10,8) & \textrm{if}\ \ 0 \leq T_sk < 15,\\
\,(10,6) & \textrm{if}\ \ 15 \leq T_sk < 30,\\
\,(10,4) & \textrm{if}\ \ 30 \leq T_sk < 45,\\
\end{cases}
\end{align*}
and Reference~2 was set to
\begin{align*}
(\bar{\theta}(k),\bar{K}_P(k))=&\ 
\begin{cases}
\,(5,9) & \textrm{if}\ \ 0 \leq T_sk < 15,\\
\,(15,6) & \textrm{if}\ \ 15 \leq T_sk < 30,\\
\,(10,7) & \textrm{if}\ \ 30 \leq T_sk < 45.\\
\end{cases}
\end{align*}
The parameters of the PI controllers in \eqref{eq:PI1} and \eqref{eq:PI2} were determined by trial and error, resulting in $G_{\rm P}^\theta=0.25$, $G_{\rm I}^\theta=0.13$, $G_{\rm P}^F=0.088$, and $G_{\rm I}^F=0.08$.
The initial voltage commands $u_1(0)=u_2(0)=5.5$ were given before the start of the control, and the control was started after sufficient time had elapsed.
Additionally, note that the simulation and experimental results regarding Reference~1 have been omitted, and the results are discussed in Section~\ref{sec:ev_q} owing to page limitations.

The simulation results are provided in Fig.~\ref{fig:sim2}, where 
(a) shows the time response of the joint angle and stiffness, 
(b) presents the tracking errors of the joint angle and stiffness, 
(c) depicts the control inputs to the valve, and 
(d) shows the inner pressure of each PAM.
Figs.~\ref{fig:sim2}\subref{sfig:ori2_out}\subref{sfig:ori2_error}  confirm that the joint angle and stiffness track the reference in a steady state.
In this case, the input voltage oscillates in a transient state, as shown in Fig.~\ref{fig:sim2}\subref{sfig:ori2_ele}. 
The operating pressure behavior is shown in Fig.~\ref{fig:sim2}\subref{sfig:ori2_pre}.
These results confirm that the proposed nonlinear controller achieves simultaneous tracking control of the joint angle and stiffness of the PAM actuator system.

However, the coefficients and input functions in the proposed controller are rational functions, which makes them difficult to implement in a controller device in an encrypted control fashion.
Therefore, in the next section, we consider polynomial approximation of the proposed controller.

\begin{remark}
The novelty of the proposed controller \eqref{eq:ctrller} is the use of the $\theta$-dependent contraction force model \eqref{eq:esti_F}, compared to the UKF-based simultaneous controller \cite{2022_shin_reference}.
The UKF-based controller involving fluid dynamics \cite{2022_shin_reference} and sliding mode controllers \cite{2011_Choi,2018_Cao} include an if-then rule; therefore, obtaining an alternative controller in a polynomial is difficult.
Meanwhile, the nonlinear controllers proposed in \cite{2009_nakamura,2016_saito,2019_Ugurl} can be applied to a polynomial approximation, whereas costly force sensors are required to measure the contraction force.
\end{remark}

\section{Polynomial Approximation of The Controller}\label{sec:polyapprox}
This section presents the approximation of the proposed controller into polynomial functions and investigates the impact of the approximation on the response of the experimental PAM control system.

\subsection{Controller Approximation}
This study uses one- and two-variable functions to approximate the reference generator \eqref{eq:barF} in polynomials because only the reference generator is a rational function.
By introducing five functions denoted by $f_1,\dots,f_5$, let the rational functions \eqref{eq:barF} be rewritten as follows:
\begin{align*}
\bar{F}_{1}(k)=&\,f_1(k;\theta,\bar{K}_P)+f_2(k;\theta)\tau_c(k)+f_3(k;\theta,P_1)+\\
&f_4(k;\theta,P_2),\\[.5ex]
\bar{F}_{2}(k)=&\,\bar{F}_{1}(k)+f_5(k;\theta)\tau_c(k),
\end{align*}
where
\begin{align*}
	f_1(k;\theta,\bar{K}_P):=&\,-\frac{l_1(k)l_2(k)\bar{K}_P(k)}{r^2(l_1(k)+l_2(k))\cos^2\theta(k)},\\
	f_2(k;\theta):=&\,\frac{l_1(k)l_2(k)}{r^2(l_1(k)+l_2(k))\cos^2\theta(k)}\Bigl(\frac{r\cos\theta(k)}{l_2(k)}-\\
	&\,\tan\theta(k)\Bigr),\\
	f_3(k;\theta,P_1):=&\,\frac{\alpha_1(P_1(k))l_2(k)}{l_1(k)+l_2(k)},\\
	f_4(k;\theta,P_2):=&\,\frac{\alpha_2(P_2(k))l_1(k)}{l_1(k)+l_2(k)},\\
	f_5(k;\theta):=&\,-\frac{1}{r\cos\theta(k)}.
\end{align*}

For the five functions, we used the LASSO-based polynomial approximation \cite{2023_Takeda} and manually removed coefficients with relatively small values, which resulted in the following approximated functions:
\begin{align*}
f_1(k;\theta,\bar{K}_P)\approx &\, w_1\bar{K}_p(k)+w_2\theta^2(k)\bar{K}_P(k)+w_3\, =: \hat{f_1},\\
f_2(k;\theta)\approx &\, w_4\theta(k)+w_5\theta^2(k)+w_6\, =: \hat{f_2},\\
f_3(k;\theta,P_1)\approx &\, w_7P_1(k)+w_8\theta(k)P_1(k)+\\
&\, w_s\theta(k)P_1^2(k)+w_9,\\
\approx &\, w_7P_1(k)+w_8\theta(k)P_1(k)+w_9\, =: \hat{f_3},\\
f_4(k;\theta,P_2)\approx &\, w_{10}P_2(k)+w_{11}\theta(k)P_2(k)+w_{12}\, =: \hat{f_4},\\
f_5(k;\theta)\approx &\, w_{13}\theta^2(k)+w_{14}\, =: \hat{f_5},
\end{align*}
where $w_i,\,\forall i\in\{1,2,\cdots,14\}$ and $w_s$ are coefficients of approximated polynomials $\hat{f}_j,\,\forall j\in\{1,2,\cdots,5\}$, and their values are listed in TABLE~\ref{tbl:keisu}, where the regularization parameter of LASSO was set to 1.0.
$\hat{f_2}$ and $\hat{f_5}$ are cubic polynomials in $\theta$, and 
the other functions are cubic in both arguments.
Furthermore, $w_s$ of $\hat{f}_3$ resulted in a nonzero term, but it was significantly small compared to the other coefficients; thus we removed $w_s$.

Consequently, we obtained the polynomial controller in \eqref{eq:ctrller} by using the approximated coefficient matrices and functions specified below,
\begin{subequations}
\label{eq:appcon}
\begin{align}
A(k;\theta)\approx &\,
\begin{bmatrix}
1 & 0 & 0\\
T_sG_\textrm{I}^\theta \hat{f_2}(k;\theta) & 1 & 0\\
T_sG_\textrm{I}^\theta(\hat{f_2}(k;\theta)+\hat{f_5}(k;\theta)) & 0 & 1
\end{bmatrix},\\[1ex]
C(k;\theta)\approx &\,
\begin{bmatrix}
G_\textrm{P}^FG_\textrm{I}^\theta \hat{f_2}(k;\theta) & G_\textrm{I}^F & 0 \\
G_\textrm{P}^FG_\textrm{I}^\theta(\hat{f_2}(k;\theta)+\hat{f_5}(k;\theta)) & 0 & G_\textrm{I}^F
\end{bmatrix},\\[1ex]
h_1(k;\zeta)\approx &\,\hat{f_1}(k;\theta,\bar{K}_P)+G_\textrm{P}^\theta \hat{f_2}(k;\theta)(\bar{\theta}(k)-\theta(k))+\notag\\
&\,\hat{f_3}(k;\theta,P_1)+\hat{f_4}(k;\theta,P_2)- \hat{a}_1P_1(k)-\hat{b}_1,\\[1ex]
h_2(k;\zeta)\approx &\,\hat{f_1}(k;\theta,\bar{K}_P)+\hat{f_3}(k;\theta,P_1)+\hat{f_4}(k;\theta,P_2)+\notag\\
&\,G_\textrm{P}^\theta(\hat{f_2}(k;\theta)+\hat{f_5}(k;\theta))(\bar{\theta}(k)-\theta(k))- \notag\\
&\,\hat{a}_2P_2(k)-\hat{b}_2.
\end{align}
\end{subequations}

\begin{table}[t]
	\centering
	\caption{List of Parameters in Approximated Controller}
	\begin{tabular}{cr|cr}\hline\hline
		$w_1$ & $-61.7$	& $p_1^{a_1}$ & 7.05$\times10^{-3}$\\
		$w_2$ & $-1.89\times 10^{-2}$ & $p_1^{a_2}$ & $-1.02\times10^{-4}$\\
		$w_3$ & $-1.58$&	$p_1^{b_1}$ & $-5.57\times10^{2}$\\
		$w_4$ & $13.6$&	$p_1^{b_2}$ & $72.86$\\
		$w_5$ & $-1.24$&	$p_2^{a_1}$ & $6.42\times10^{3}$\\
		$w_6$ & $1.73\times10^{-3}$ & $p_2^{a_2}$ & $-9.18\times10^{-4}$\\ 
        $w_7$ & $-5.12\times10^{-1}$&  $p_2^{b_1}$ & $-1.98\times10^{2}$\\ 
		$w_8$ & $-1.18\times10^{-3}$ & $p_2^{b_2}$ & 15.75\\ 
		$w_s$ & $-5.76\times10^{-7}$ & $p_1^{\hat{a}_1}$ & $-2.55\times10^{-4}$\\
		$w_9$ & 36.5 & $p_1^{\hat{a}_2}$ & $1.45\times10^{-4}$\\	
        $w_{10}$ & $-4.31\times10^{-1}$ &  $p_1^{\hat{b}_1}$ & 20.14\\ 
		$w_{11}$ & $1.54\times10^{-3}$ & $p_1^{\hat{b}_2}$ & $-19.01$\\ 
		$w_{12}$ & $-9.35$& $p_2^{\hat{a}_1}$ & $2.18\times10^{-4}$\\
		$w_{13}$ & $-27.3$ & $p_2^{\hat{a}_2}$ & $1.45\times10^{-4}$\\
		$w_{14}$ & $-3.92\times10^{-3}$ &	$p_2^{\hat{b}_1}$ & $1.35$\\ 
		$G_\textrm{P}^\theta$ & 1.3 & $p_2^{\hat{b}_2}$ & $-18.67$\\
		$G_\textrm{I}^\theta$& $2.43\times10^{-1}$ & $G_\textrm{I}^F$& $2.50\times10^{-2}$ \\ 
		$G_\textrm{P}^F$& $8.80\times10^{-2}$& &  \\ 
		\hline
		\hline
	\end{tabular}
	\label{tbl:keisu}
\end{table}

\subsection{Evaluation of Approximation Impacts}
The impact of the controller approximation on the control performance is evaluated by comparing the experimental time responses of the control systems using the original controller \eqref{eq:ACgh} and the approximated controller \eqref{eq:appcon}.
The scenarios of the control experiments are simultaneous tracking control of the joint angle and stiffness with and without a load.
The references of the joint angle and stiffness set in the experiments are the same as those employed in the simulations.
Before the start of the control, an initial voltage command $u_1(0)=u_2(0)=5.5$ were provided, and the control was initiated after sufficient time had elapsed.

\begin{figure*}[htbp]
\centering
	\subfloat[]{\includegraphics[scale=.54]{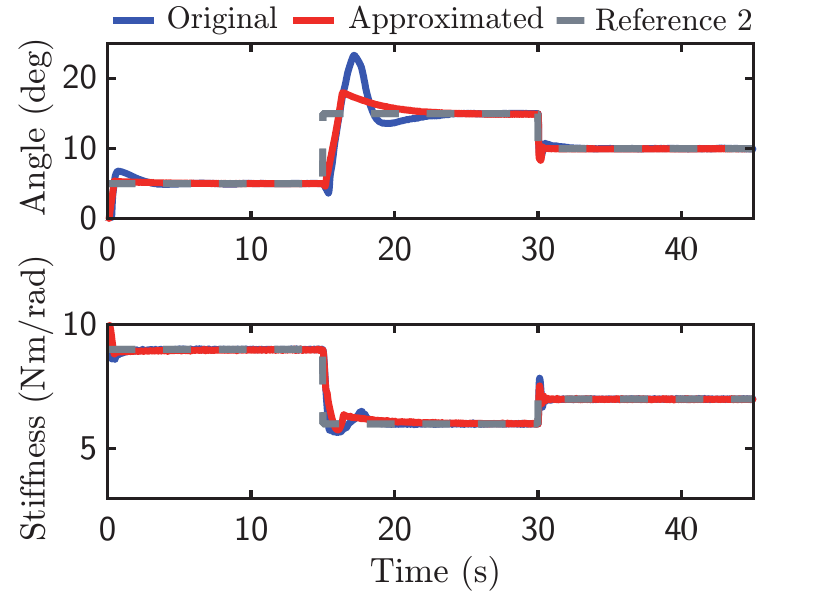}\label{sfig:app2_out}}
	\subfloat[]{\includegraphics[scale=.54]{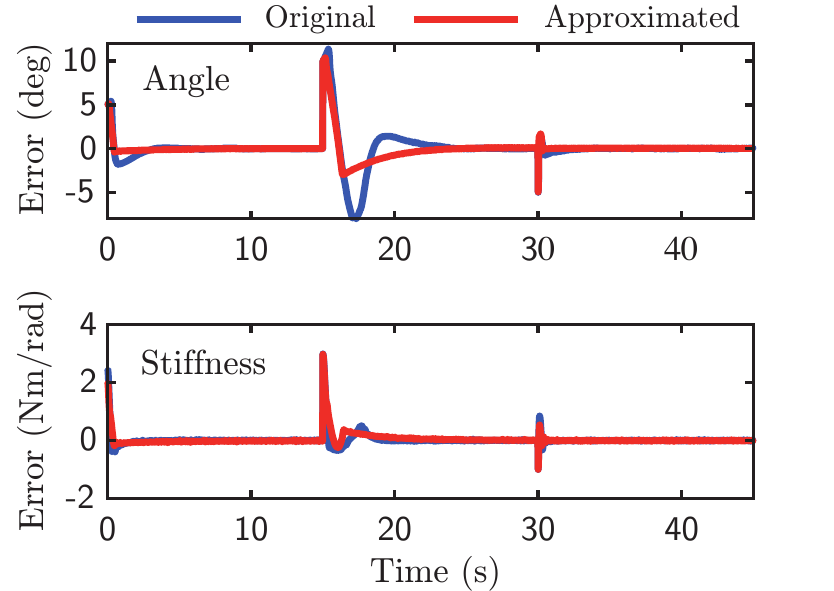}\label{sfig:app2_error}}
    \subfloat[]{\includegraphics[scale=.54]{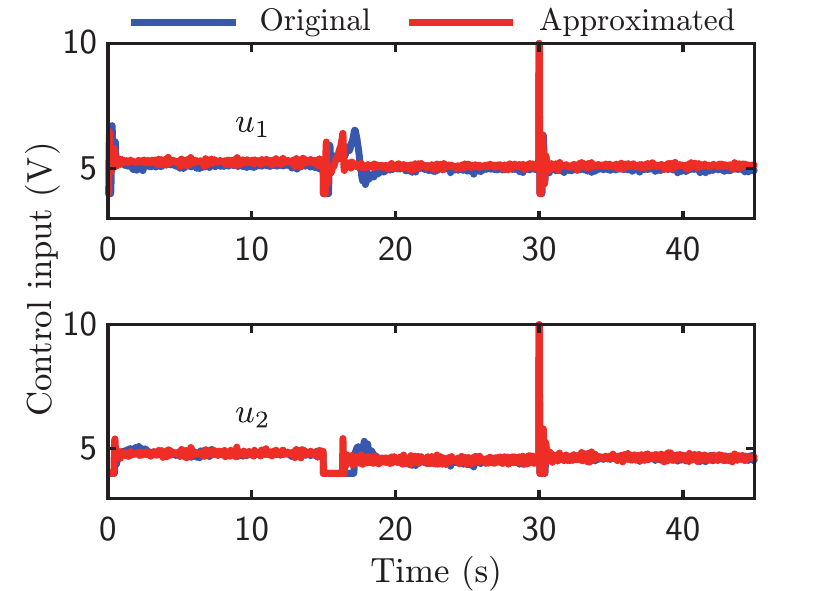}\label{sfig:app2_ele}}
	\subfloat[]{\includegraphics[scale=.54]{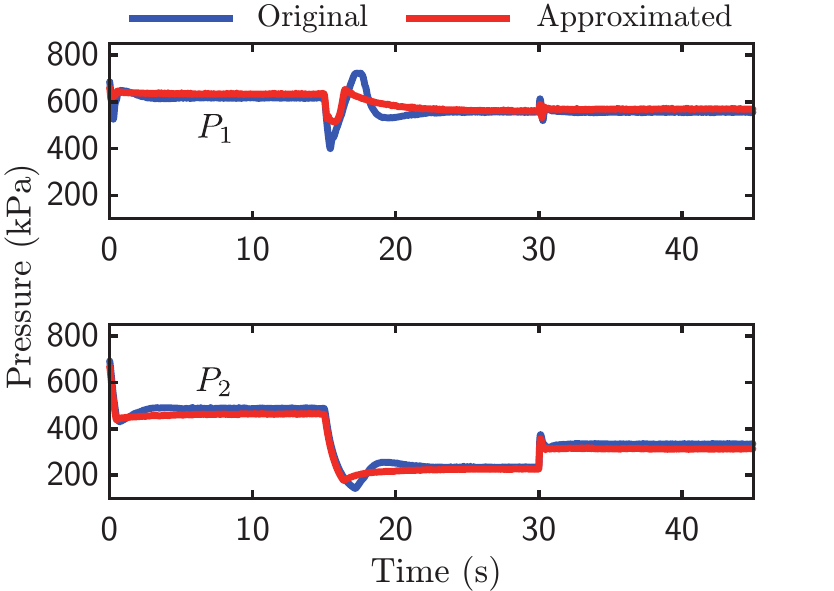}\label{sfig:app2_pre}}
    \caption{Comparison of experimental results of original and approximated polynomial load-free controls of joint angle and stiffness for Reference~2.
    (a) Joint angle and stiffness, (b) tracking errors of the joint angle and stiffness, (c) control inputs to the valve, and (d) inner pressure of each PAM.}
\label{fig:exp2_wol}
\end{figure*}

The experimental results for the original and approximated controllers without a load are shown in Fig.~\ref{fig:exp2_wol}.
In this figure, the blue and red lines indicate the results of the original and approximated control systems, respectively, and their meanings are the same as those in Fig.~\ref{fig:sim2}.
Figs.~\ref{fig:exp2_wol}\subref{sfig:app2_out}\subref{sfig:app2_error} confirm that the angle and stiffness track the reference in the steady state and that sufficient control performance is maintained.
Additionally, the improvement in transient-response control performance is believed to be due to the impacts of quantization; however, the details have not been clearly established.
Figs.~\ref{fig:exp2_wol}\subref{sfig:app2_ele}\subref{sfig:app2_pre} observe that the experimental results of the original and approximated controllers exhibit similar behavior in the steady state.

\begin{figure*}[htbp]
\centering
	\subfloat[]{\includegraphics[scale=.54]{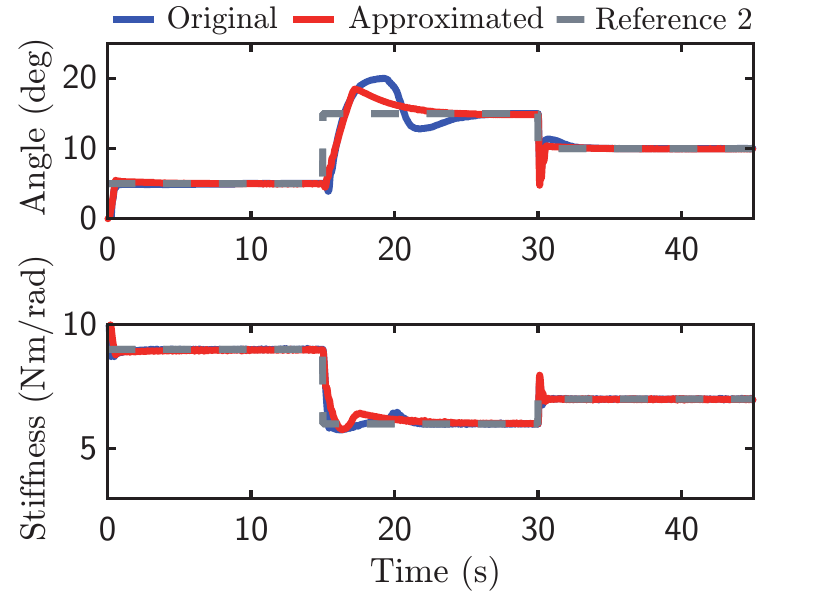}\label{sfig:app4_out}}
	\subfloat[]{\includegraphics[scale=.54]{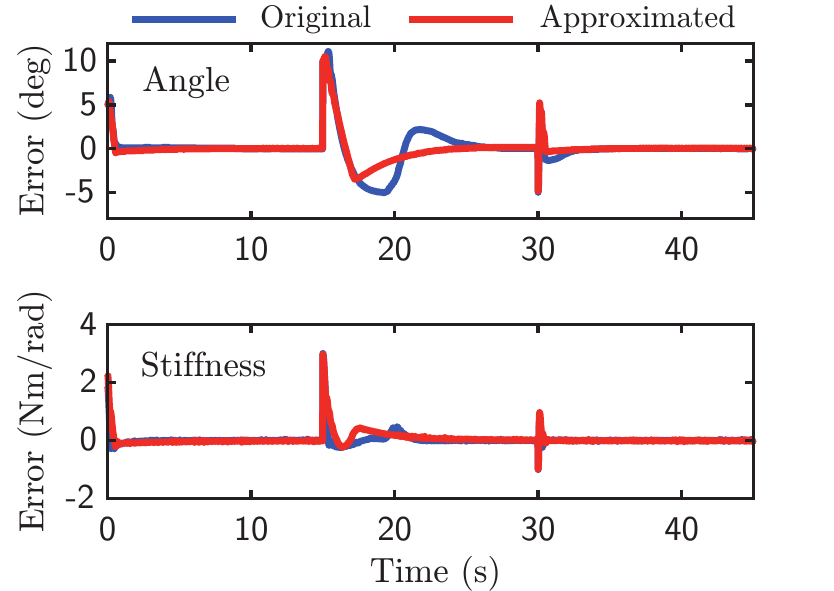}\label{sfig:app4_error}}
	\subfloat[]{\includegraphics[scale=.54]{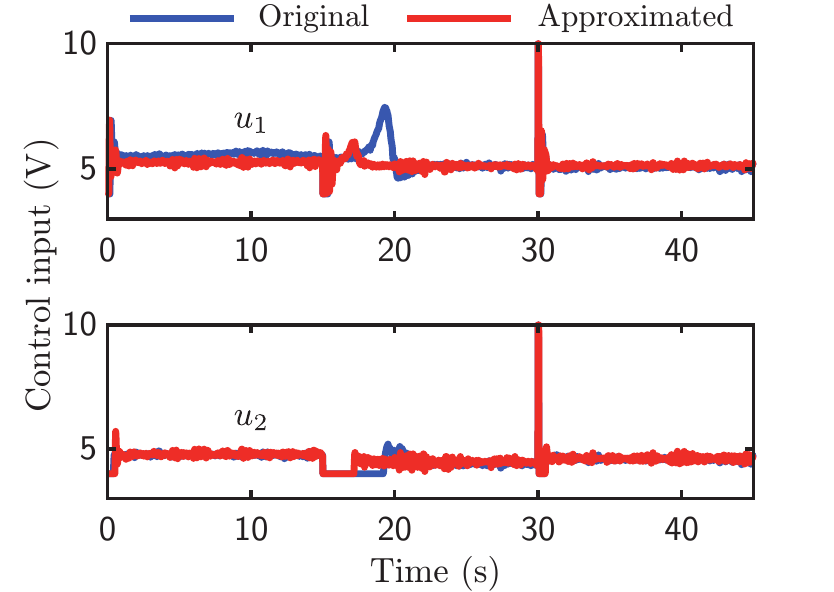}\label{sfig:app4_ele}}
	\subfloat[]{\includegraphics[scale=.54]{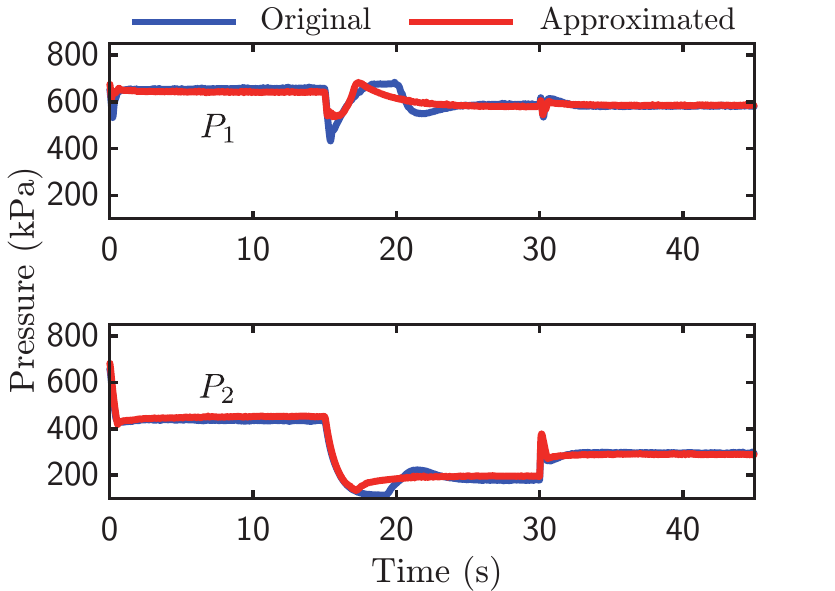}\label{sfig:app4_pre}}
\caption{Comparison of experimental results of original and approximated polynomial controls with load for Reference~2. (a) Joint angle and stiffness, (b) Tracking errors of the joint angle and stiffness, (c) Control inputs to the valve, and (d) Inner pressure of each PAM.}
\label{fig:exp2_wl}
\end{figure*}

Moreover, we conducted a control experiment in which a 1.5 kg load was hung from the left side of the joint before starting control to evaluate the impact of the approximation.
The experimental results of the loads are shown in Fig.~\ref{fig:exp2_wl}, where the line colors in each figure are the same as those in Fig.~\ref{fig:exp2_wol}.
Figs.~\ref{fig:exp2_wl}\subref{sfig:app4_out}\subref{sfig:app4_error} confirm that the angle and stiffness track the reference in the steady state.
Good control tracking causes the controller to compensate for the impact of the load, which can be observed as a difference of approximately 50 kPa over the steps between the responses of the inner pressures, as shown in Figs.~\ref{fig:exp2_wol}\subref{sfig:app2_pre} and \ref{fig:exp2_wl}\subref{sfig:app4_pre}.
Similarly, Fig.~\ref{fig:exp2_wl}\subref{sfig:app4_ele} demonstrates that the experimental results of the two controllers exhibit similar behaviors in the steady state, respectively. 

The experimental results confirm that the approximated controller achieves almost the same control performance as the original controller, implying that the impact of the controller approximation is negligible.
Therefore, the original controller can be replaced with an approximated controller that is adequate for secure implementation.

\section{Secure PAM Actuator with Encrypted Polynomial-Type Controller}\label{sec:ec}
This section describes the encrypted control of the joint angle and the stiffness of an antagonistic PAM actuator system. 

\subsection{Secure Implementation}
\begin{figure}[tb]
	\centering
	\includegraphics[keepaspectratio,width=.95\hsize]{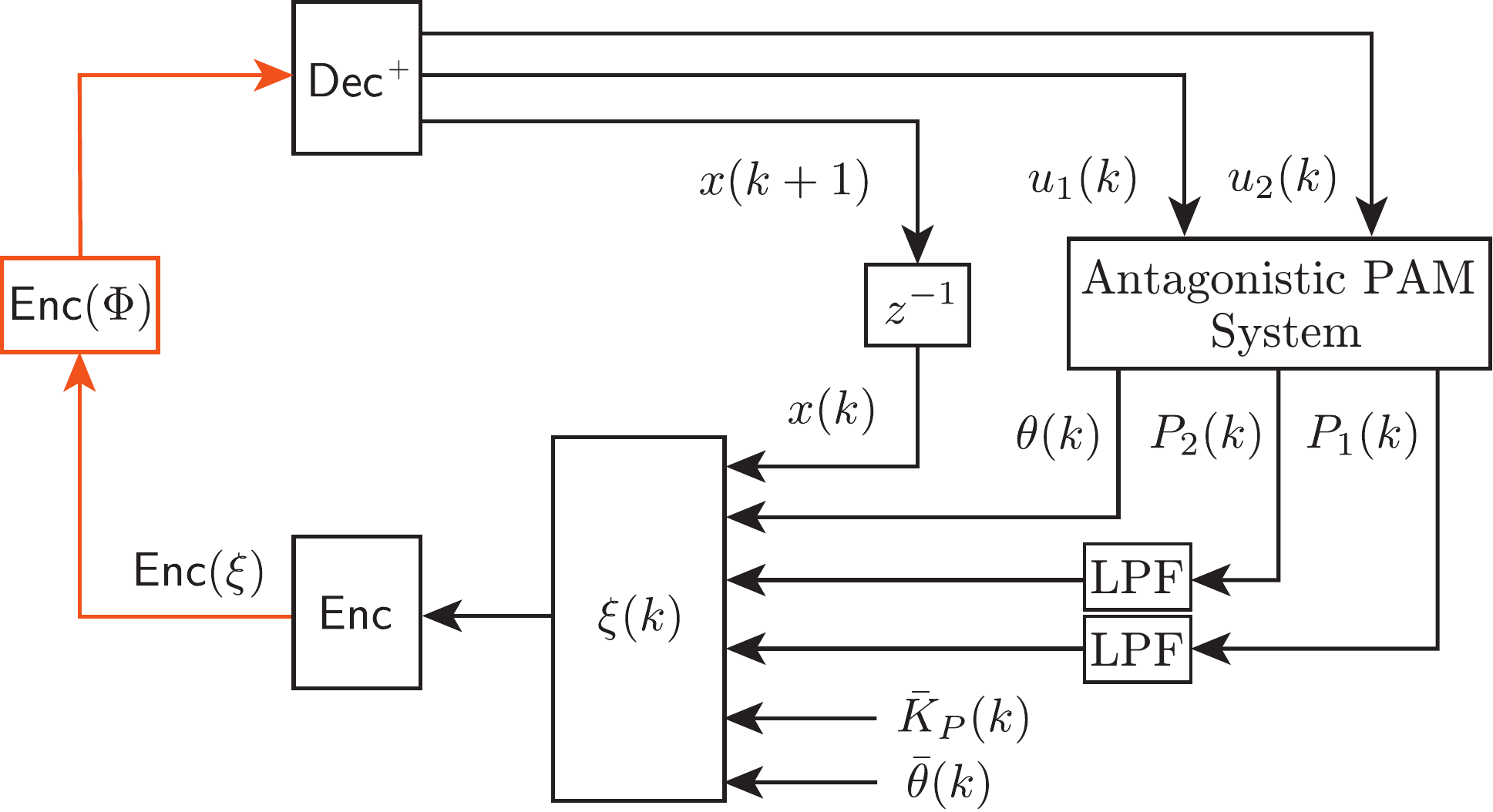}\vspace{3mm}
	\caption{Encrypted angle and stiffness control system.}
	\label{fig:enc_controller}
\end{figure}

The controller encryption method in~\cite{2015_Kogiso} can be applied to a linear controller in a matrix-vector product form.
Hence, we consider to transform our proposed polynomial controller into a product of matrix and vector.
In this study, by defining a vector of monomials as $\xi\in\R^{18}$, the polynomial controller~\eqref{eq:ctrller} with the coefficients and functions~\eqref{eq:appcon} can be transformed into the following matrix-vector multiplication:
\begin{align}
    \psi(k)=&\,\Phi\xi(k),\ \,\forall k\in\Z^+, \label{eq:pcon}
\end{align}
with 
\begin{align*}
 \xi:=&\, [\,\bar{K}_P\,\ \bar{K}_P\theta^2\,\ \bar{\theta}\,\ \theta\,\ x^\theta\,\ \bar{\theta}\theta\,\ \theta^2\,\ x^\theta\theta\,\ \bar{\theta}\theta^2\,\ \theta^3\,\ x^\theta\theta^2\,\ P_1\\
	&\ \ \theta P_1\,\ P_2\,\ \theta P_2\,\ 1\,\ x^{F_1}\,\ x^{F_2}\,]^\mathrm{T},\\
	\psi(k):=&\, [\,x^\theta(k+1)\,\ x^F_1(k+1)\,\ x^F_2(k+1)\,\ u_1(k)\,\ u_2(k)\,]^\mathrm{T},
\end{align*}
where $\xi$ and $\psi\in\R^5$ are the input and output of the encrypted controller, respectively.
The coefficient matrix $\Phi\in\R^{5\times18}$ is given as follows:
\begin{align*}
	\setcounter{MaxMatrixCols}{18}
	\Phi = 
	\left[
		\begin{array}{cccccc}
		0&0&T_s&-T_s&1&0\\
    T_sw_1&T_sw_2&T_s\phi_3&T_s\phi_4&T_s\phi_5&T_s\phi_6\\
		T_sw_1&T_sw_2&T_s\nu_3&T_s\nu_4&T_s\nu_5&T_s\nu_6\\
		G_P^Fw_1&G_P^Fw_2&G_P^F\phi_3&G_P^F\phi_4&G_P^F\phi_5&G_P^F\phi_6\\
		G_P^Fw_1&G_P^Fw_2&G_P^F\nu_3&G_P^F\nu_4&G_P^F\nu_5&G_P^F\nu_6\\
		\end{array}\right.\nonumber\\
		\left. 
		\begin{array}{cccccc}
		0&0&0&0&0&0\\
		T_s\phi_7&T_s\phi_8&T_s\phi_9&T_s\phi_{10}&T_s\phi_{11}&T_s\phi_{12}\\
		G_P^F\phi_7&G_P^F\phi_8&G_P^F\phi_9&G_P^F\phi_{10}&G_P^F\phi_{11}&T_s\nu_{12}\\
		G_P^F\phi_7&G_P^F\phi_8&G_P^F\phi_9&G_P^F\phi_{10}&G_P^F\phi_{11}&G_P^F\phi_{12}\\
		G_P^F\nu_7&G_P^F\nu_8&G_P^F\nu_9&G_P^F\nu_{10}&G_P^F\nu_{11}&G_P^F\nu_{12}\\
		\end{array}\right.\nonumber\\
		\left.
		\begin{array}{cccccc}
		0&0&0&0&0&0\\
		T_s\phi_{13}&T_s\phi_{14}&T_s\phi_{15}&T_s\phi_{16}&1&0\\
		T_s\nu_{13}&T_s\nu_{14}&T_s\nu_{15}&T_s\nu_{16}&0&1\\
		G_P^F\phi_{13}&G_P^F\phi_{14}&G_P^F\phi_{15}&G_P^F\phi_{16}+\beta_1&G_I^{F}&0\\
		G_P^F\nu_{13}&G_P^F\nu_{14}&G_P^F\nu_{15}&G_P^F\nu_{16}+\beta_2&0&G_I^{F}
		\end{array}
	\right],
\end{align*}
where 
$\phi_3:=w_4G_P^\theta$, 
$\phi_4:=-p_1^{b_1}-w_4G_P^\theta$, 
$\phi_5:=w_4G_I^\theta$, 
$\phi_6:=w_5G_P^\theta$, 
$\phi_7:=-w_5G_P^\theta$, 
$\phi_8:=w_5G_I^\theta$, 
$\phi_9:=w_6G_P^\theta$, 
$\phi_{10}:=-w_6G_P^\theta$, 
$\phi_{11}:=w_6G_I^\theta$, 
$\phi_{12}:=w_7-p_2^{a_1}$, 
$\phi_{13}:=w_8-p_1^{\hat{a}_1}$, 
$\phi_{14}:=w_{10}$, 
$\phi_{15}:=w_{11}$, 
$\phi_{16}:=w_3+w_9+w_{12}-p_2^{\hat{b}_1}$, 
$\nu_3:=(w_4+w_{13})G_P^\theta$, 
$\nu_4:=-p_1^{\hat{b}_2}-(w_4+w_{13})G_P^\theta$, 
$\nu_5:=(w_4+w_{13})G_I^\theta$, 
$\nu_6:=w_5G_P^\theta$, 
$\nu_7:=-w_5G_P^\theta$, 
$\nu_8:=w_5G_I^\theta$, 
$\nu_9:=(w_6+w_{14})G_P^\theta$, 
$\nu_{10}:=-(w_6+w_{14})G_P^\theta$, 
$\nu_{11}:=(w_6+w_{14})G_I^\theta$, 
$\nu_{12}:=w_7$, 
$\nu_{13}:=w_8$, 
$\nu_{14}:=w_{10}-p_2^{\hat{a}_2}$, 
$\nu_{15}:=w_{11}-p_1^{\hat{a}_2}$, and 
$\nu_{16}:=w_3+w_9+w_{12}-p_2^{\hat{b}_2}$.
The augmented formulation of the controller \eqref{eq:pcon} is a linear operation; thus, the same procedure for encrypting a linear controller is applied to encrypt the controller, as described in \cite{2022_shin_cyber}.
The configuration of the control system is illustrated in~Fig.\,\ref{fig:enc_controller}, where the controller input $\xi$ is generated by the reference signals, measurements, and states of the PI controllers.
The input $\xi$ is encrypted by using an ElGamal encryption scheme in the \textsf{Enc} block.
The encrypted controller $\Enc(\Phi)$ conducts multiplicative homomorphic operations, and $\textsf{Dec}^+$ extracts the plaintext states and control inputs to each PDCV.
For the notations regarding the encryption scheme, please refer to \cite{2015_Kogiso}.

\subsection{Experimental Validation}
This section verifies the encrypted control system by comparing the experimental results obtained using the encrypted and original control systems.
A key length of 64~bit was chosen, and the scaling parameters $\Delta_\xi$ and $\Delta_\Phi$ were set to $1.0\times10^8$, which were introduced in \cite{2015_Kogiso} to manage the quantization errors due to encryption.

\begin{figure*}[htbp]
\centering
	\subfloat[]{\includegraphics[scale=.54]{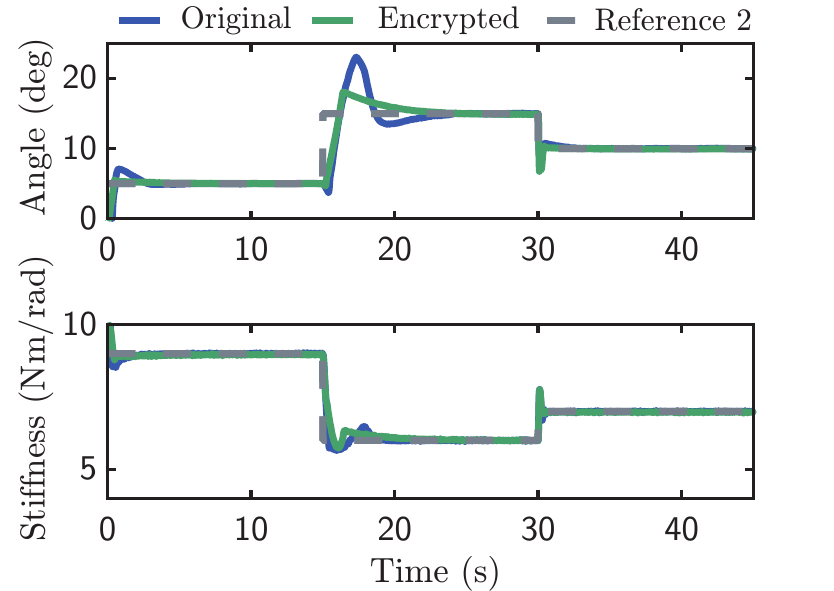}\label{sfig:enc2_out}}
	\subfloat[]{\includegraphics[scale=.54]{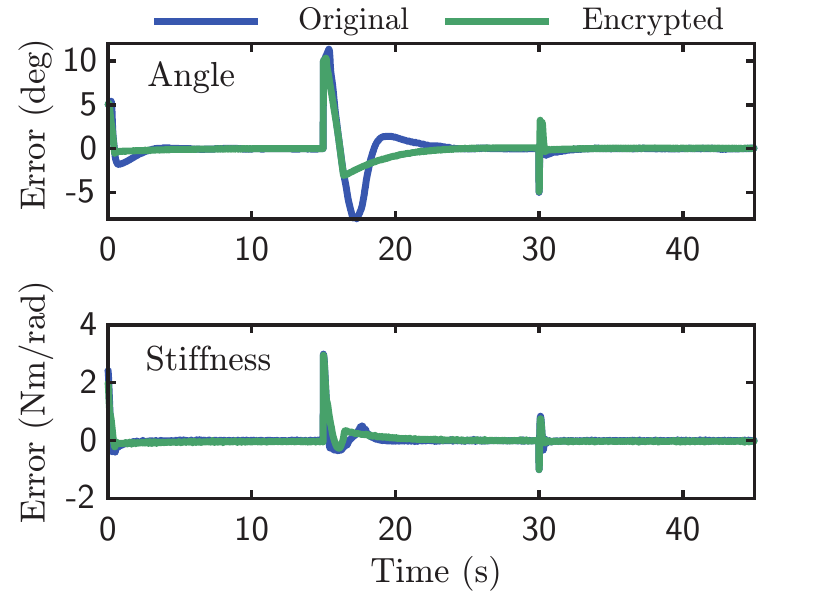}\label{sfig:enc2_error}}
    \subfloat[]{\includegraphics[scale=.54]{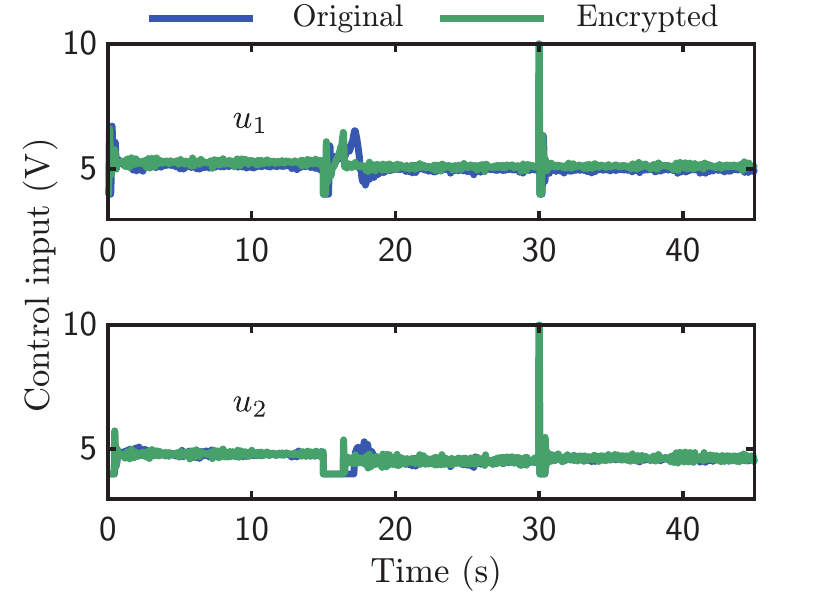}\label{sfig:enc2_ele}}
	\subfloat[]{\includegraphics[scale=.54]{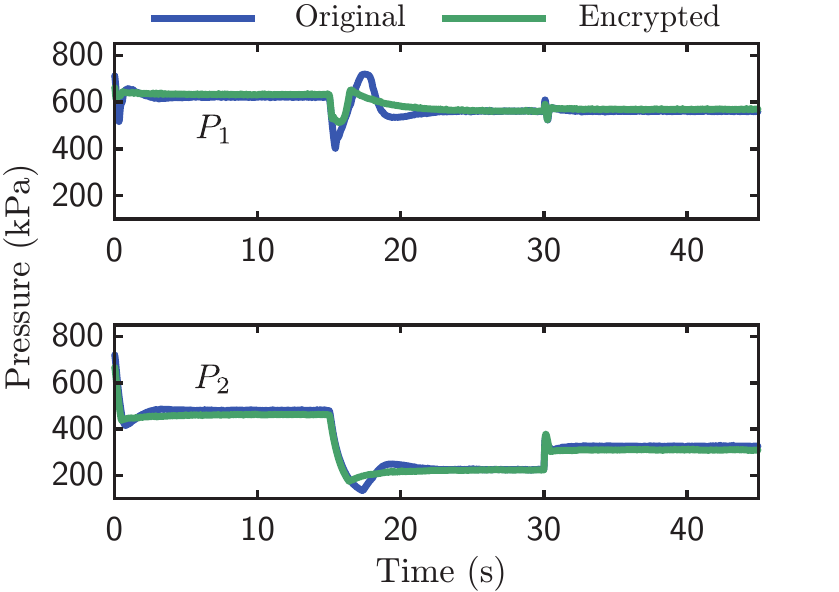}\label{sfig:enc2_pre}}
    \caption{Experimental results of unencrypted and encrypted simultaneous load-free control of joint angle and stiffness for Reference~2. (a) Joint angle and stiffness, (b) tracking errors of the joint angle and stiffness, (c) control inputs to the valve, and (d) inner pressure of each PAM.}
\label{fig:exp4_wol}
\end{figure*}

\begin{figure*}[htbp]
\centering
	\subfloat[]{\includegraphics[scale=.54]{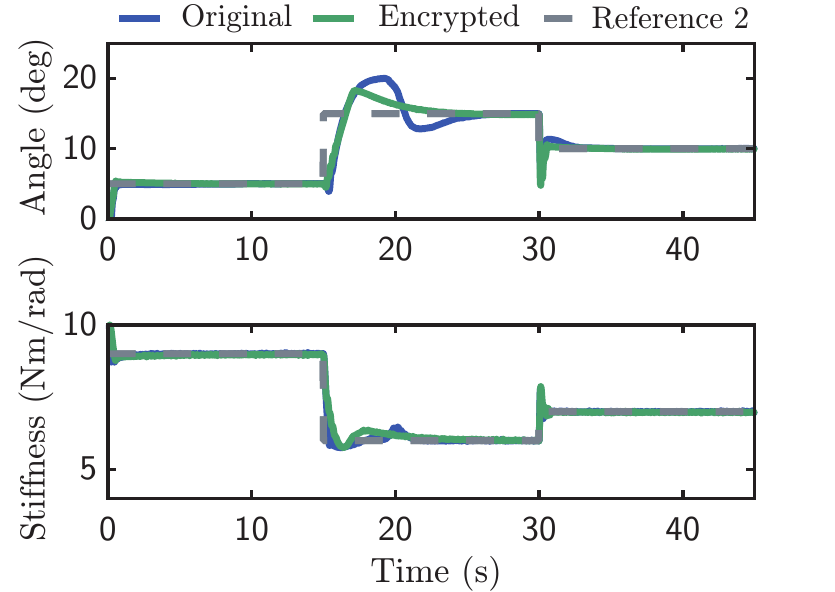}\label{sfig:enc4_out}}
	\subfloat[]{\includegraphics[scale=.54]{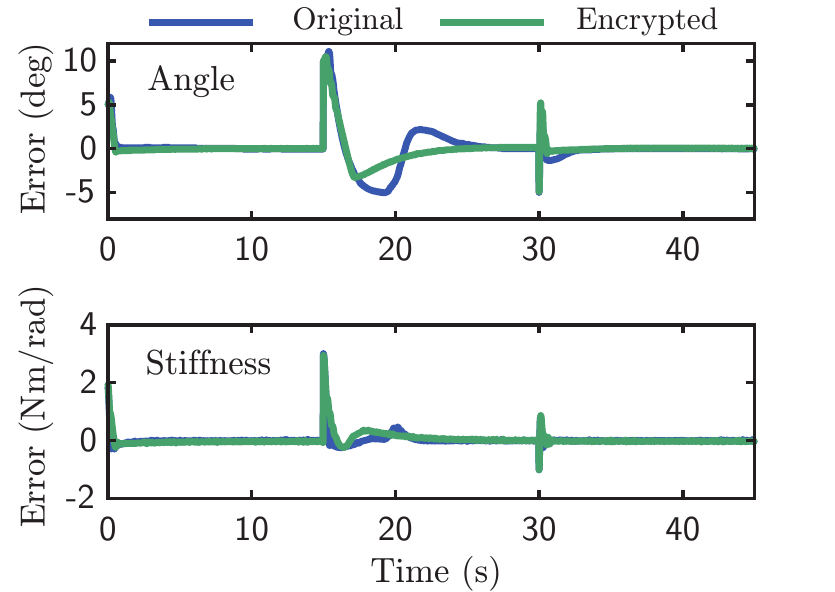}\label{sfig:enc4_error}}
    \subfloat[]{\includegraphics[scale=.54]{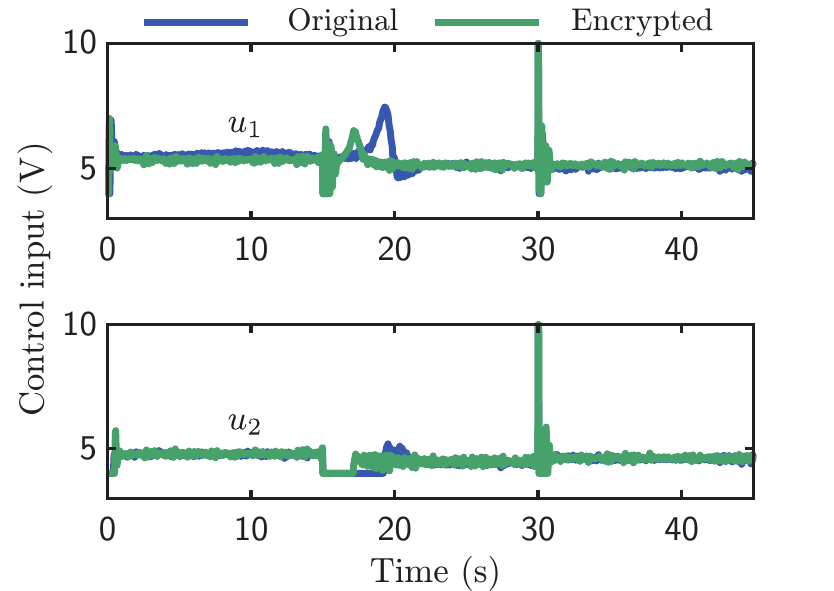}\label{sfig:enc4_ele}}
	\subfloat[]{\includegraphics[scale=.54]{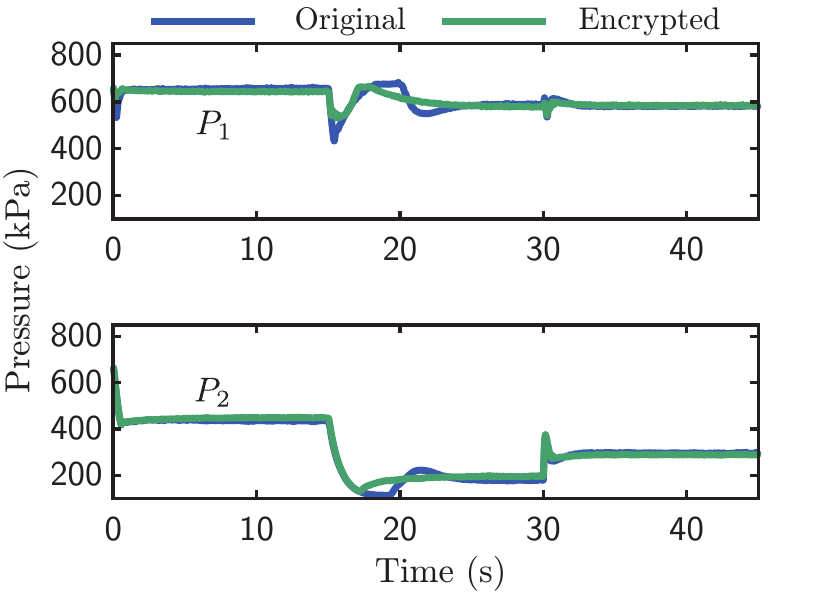}\label{sfig:enc4_pre}}
    \caption{Experimental results of unencrypted and encrypted simultaneous controls with load for Reference~2. (a) Joint angle and stiffness, (b) tracking errors of the joint angle and stiffness, (c) control inputs to the valve, and (d) inner pressure of each PAM.}
\label{fig:exp4_wl}
\end{figure*}

The scenarios of the control experiments involve simultaneous tracking control of the joint angle and stiffness with and without the load.
The references of the joint angle and stiffness were the same as those of the simulations in Figs.~\ref{fig:sim2}.
Before the start of the control, an initial voltage command $u_1(0)=u_2(0)=5.5$ was provided, and the control was initiated after sufficient time had elapsed.
The experimental results are shown in Figs.~\ref{fig:exp4_wl}.
In the figures, the blue and green lines represnet the results of the original and encrypted control systems, respectively.
Figs.~\ref{fig:exp4_wol}\subref{sfig:enc2_out}\subref{sfig:enc2_error} and \ref{fig:exp4_wl}\subref{sfig:enc4_out}\subref{sfig:enc4_error} confirm that the angle and stiffness track the reference in the steady state and that sufficient control performance is maintained.
Similarly,they show that the controller compensated for the impact of the load, which can be observed as a difference of approximately 50~kPa.

\begin{figure}[t]
	\centering
 	\subfloat[]{\includegraphics[scale=.54]{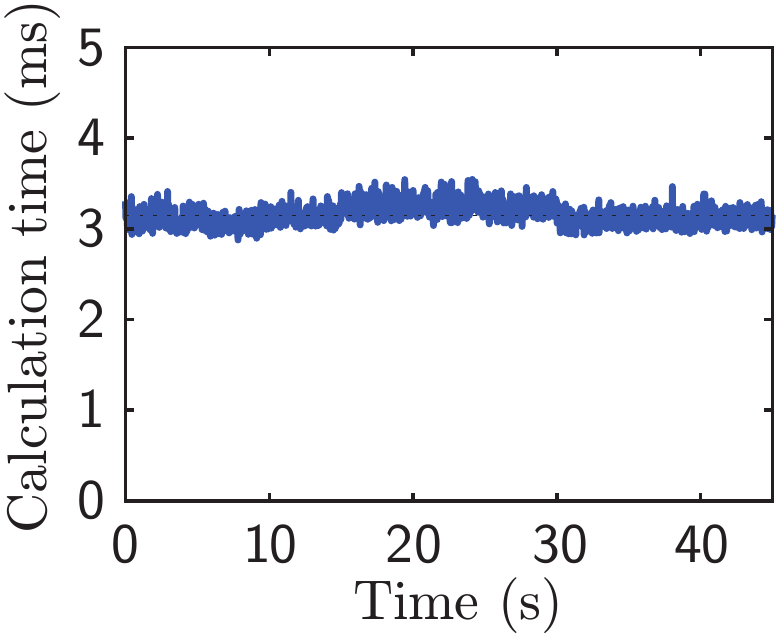}	\label{fig:cal_time_b}}
   	\subfloat[]{\includegraphics[scale=.54]{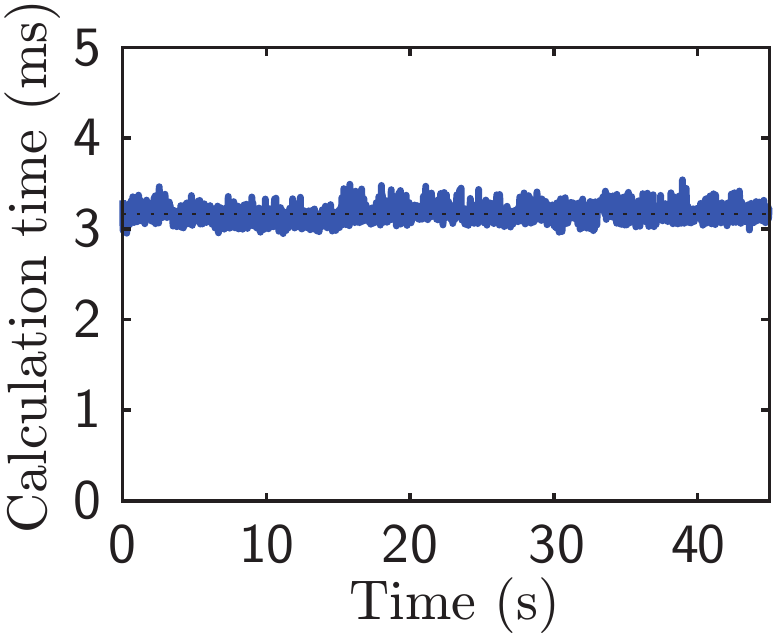}	\label{fig:cal_time_d}}
	\caption{Processing time: less than the sampling period of~20 ms. (a) Reference~2 without load and (b) Reference~2 with load.}
	\label{fig:cal_time}
\end{figure}

The computation time involved in the encrypted control system is shown in Fig.~\ref{fig:cal_time}, where Figs.~\ref{fig:cal_time}\subref{fig:cal_time_b}\subref{fig:cal_time_d} correspond to Figs.~\ref{fig:exp4_wol} and \ref{fig:exp4_wl}.
The figures show that the control operation is in real time because the computation time for each step is less than the sampling period of 20~ms.

\subsection{Quantitative Investigation}\label{sec:ev_q}
We quantitatively evaluate the control results presented so far to investigate the impact of the secure implementation of the nonlinear controller on the control performance.
For this purpose, we introduce the $\ell_2$ norm to measure the tracking error during a specific duration for each control result, as shown in Figs.~\ref{fig:exp2_wol}-\ref{fig:exp2_wl} and \ref{fig:exp4_wol}-\ref{fig:exp4_wl}.
The $\ell_2$ norm is defined as 
$\gamma(z,\bar{z}):=\sqrt{\sum_{k=k_0}^{k_1}(z(k)-\bar{z}(k))^2}$, 
where $z$ and $\bar{z}$ represent scalar signal sequences, and the evaluation steps (closed interval) $[k_0,k_1]$ are set to $[500,749]$, $[1250,1499]$, and $[2000,2249]$.
These intervals are labeled \#1, \#2, and \#3, respectively, in References 1 and 2.

\begin{figure*}[tbp]
	\centering
	\subfloat[]{\includegraphics[scale=.55]{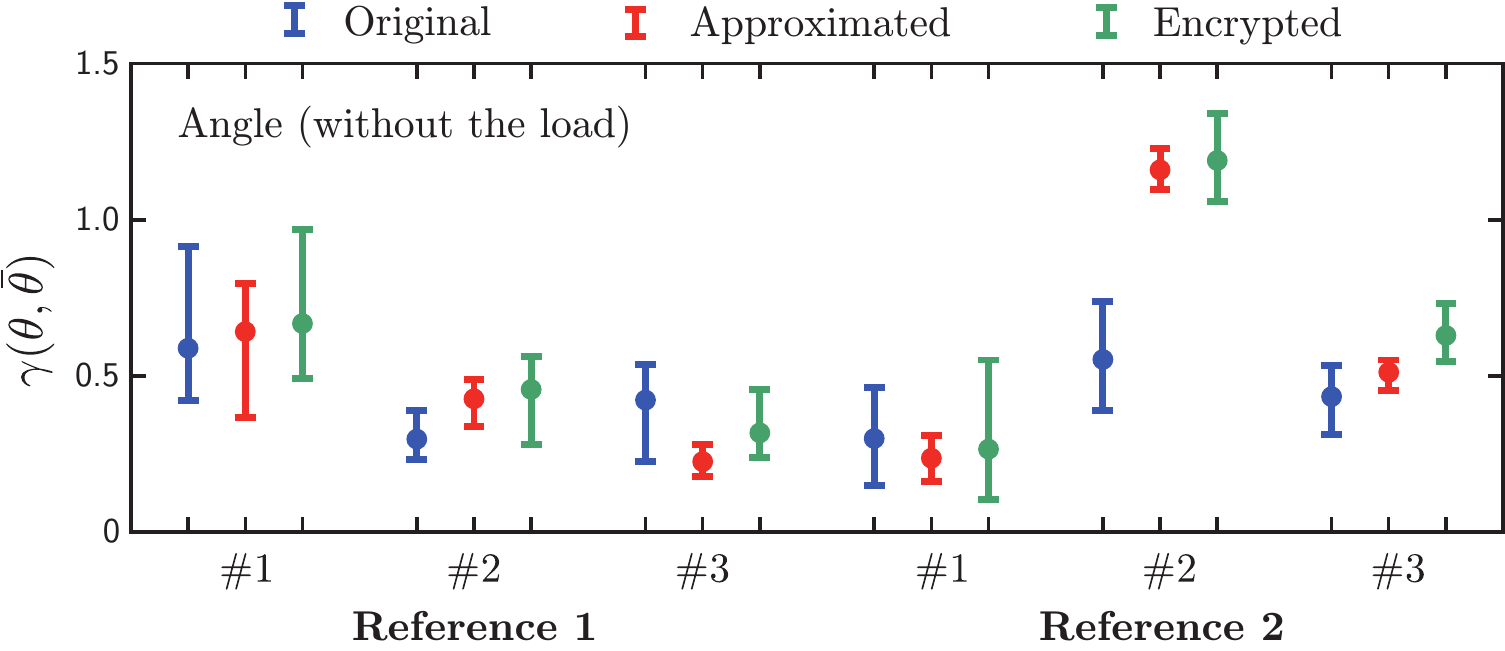}\label{sfig:con1_angle}}\ \ \ 
	\subfloat[]{\includegraphics[scale=.55]{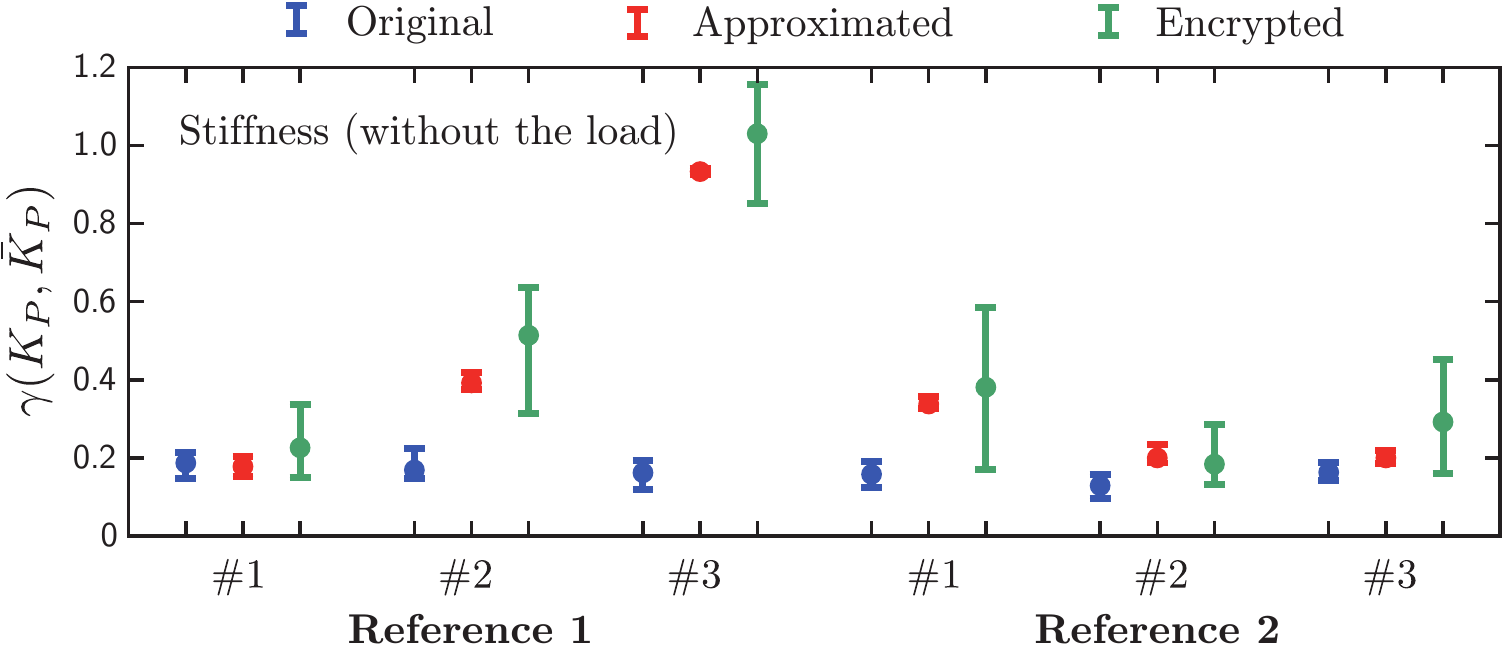}\label{sfig:con1_stiff}}
	\caption{$\ell_2$-norm scores of the tracking errors of the (a) joint angle and (b) stiffness using the original, approximated, and encrypted controllers without a load.}\label{fig:com}
\end{figure*}

\begin{figure*}[tbp]
	\centering
	\subfloat[]{\includegraphics[scale=.55]{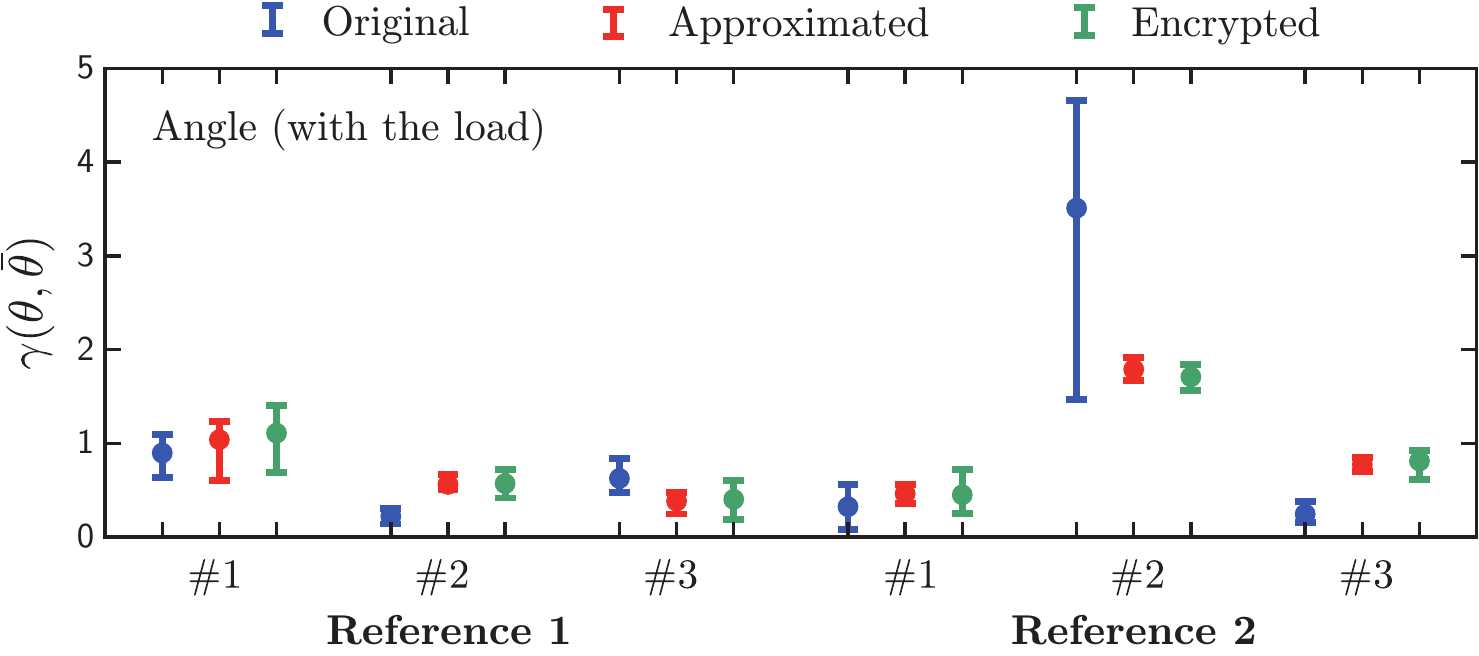}\label{sfig:con2_angle_load}}\ \ \ 
	\subfloat[]{\includegraphics[scale=.55]{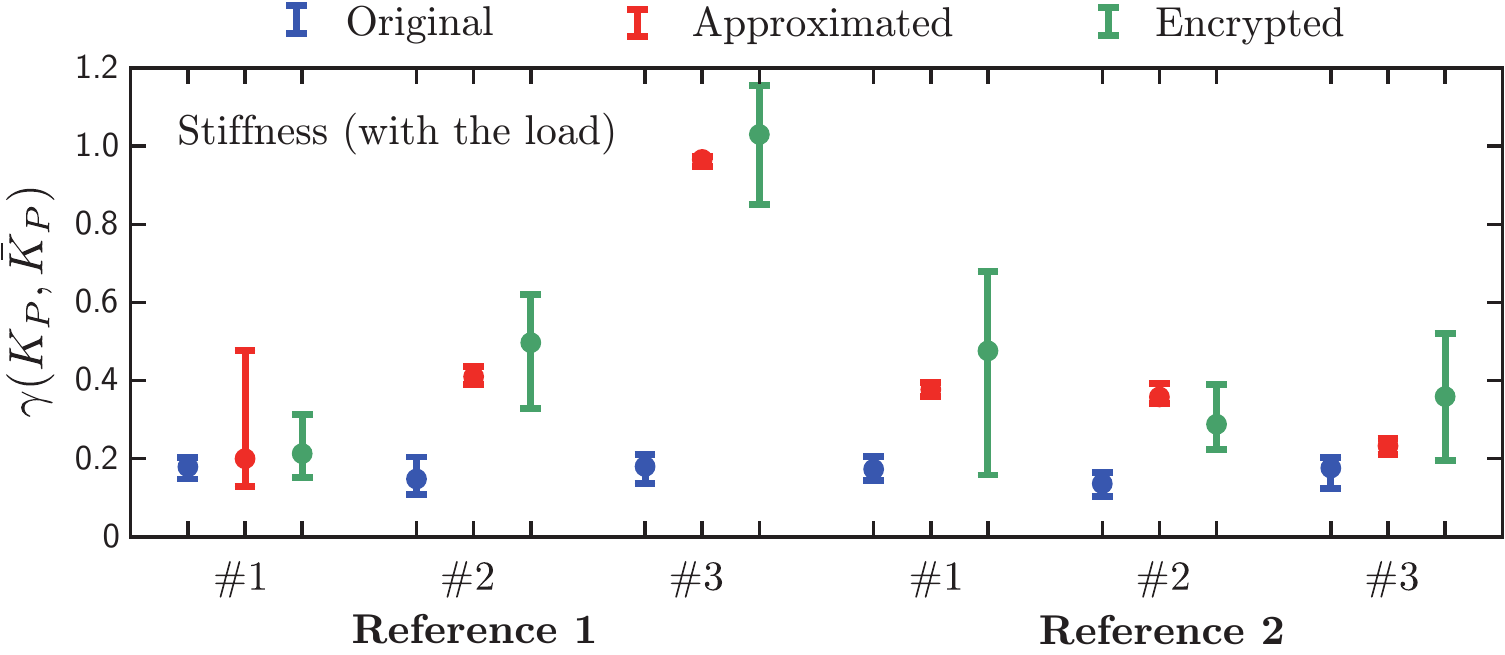}\label{sfig:con2_stiff_load}}
	\caption{$\ell_2$-norm scores of the tracking error of the (a) joint angle and (b) stiffness using the original, approximated, and encrypted controllers with a load.}\label{fig:com_load}
\end{figure*}

The $\ell_2$-norm scores of the original, approximated, and encrypted controls with and without loads are summarized in Figs.~\ref{fig:com} and \ref{fig:com_load}, respectively.
In each figure, \subref{sfig:con1_angle} and \subref{sfig:con1_stiff} display $\gamma(\theta,\bar\theta)$ and $\gamma(K_P,\bar{K}_P)$, respectively, for the three evaluation steps. 
The blue, red, and green colors represent the $\ell_2$-norm values of the original, approximated, and encrypted controllers, respectively.
The dots and error bars indicate the average, maximum, and minimum $\ell_2$-norm values from 10 experiments.

These figures confirm that the resulting scores tend to increase as the procedure advances through the polynomial approximation and secure implementation.
However, in many cases, the scores of the original and encrypted controllers are similar.
The worst rate of the controlled signal relative to its reference is $3.89/4.00\times$ 100=97.3~\% for \#3 of Reference 1 in Fig.~\ref{fig:com}\subref{sfig:con1_stiff}.
This result implies that the proposed encrypted control achieved a tracking error of less than 2.7~\%. 
Moreover, in cases where a relatively large change in the score occurs, such as \#2 of Reference 2 in Fig.~\ref{fig:com}\subref{sfig:con1_angle} and \#3 of Reference 1 in Fig.~\ref{fig:com_load}\subref{sfig:con2_stiff_load}, the effect of the polynomial approximation prevails over the change in.
In other words, the change in score from blue to red is larger than that from red to green.
This discussion provides further insight, suggesting that increasing the accuracy of the polynomial approximation could help avoid the degradation of the original control performance.
It is crucial to consider an accuracy-aware approximation method suitable for secure implementation, which will be the focus of future work.
In addition, these figures confirm that, owing to the lack of significant differences between the cases without and with the load, the proposed control system is capable of compensating for the unknown load.

\section{Conclusion}\label{sec:con}
This study proposed an encrypted simultaneous control system for antagonistic PAM actuators aimed at developing cyber-secure and safe PAM actuator systems. 
A novel nonlinear controller was designed to track the joint angle and stiffness simultaneously based on the PAM actuator model. 
By applying the polynomial approximation technique to a nonlinear controller, we obtained a polynomial-type controller.
Subsequently, through the secure implementation in the control device, we developed a secure PAM actuator system.
For experimental validation, the $\ell_2$ norm was introduced to measure and compare the experimental results of the original, approximated, and encrypted controllers. 
The experimental results showed that the proposed encrypted controller achieved simultaneous tracking of the joint angle and stiffness of the PAM actuator with a tracking error of less than 2.7~\%. 
Consequently, the developed PAM actuator system, enabled by the secure implementation of the simultaneous controller, enhances security while maintaining a control performance similar to that of the original controller.
The developed actuator is expected to be used in secure and safe PAM-driven devices, such as nursing care robots, rehabilitation orthoses, and power-assisted orthoses for remote usage applications.

In future work, we plan to improve the control performance of the encrypted controller further by addressing the nonlinear characteristics of PAMs, including Coulomb friction and fluid dynamics. 
To mitigate the performance degradation caused by the polynomial approximation, we investigate more effective strategies for tuning the regularization parameter in LASSO. 
Moreover, when considering the secure implementation of cost-effective computers, reducing the computation time is essential to achieve a resource-aware encrypted controller by streamlining the proposed controller.

\bibliographystyle{IEEEtran}
\bibliography{reference}

\begin{IEEEbiography}[{\includegraphics[width=1in,height=1.25in,clip,keepaspectratio]{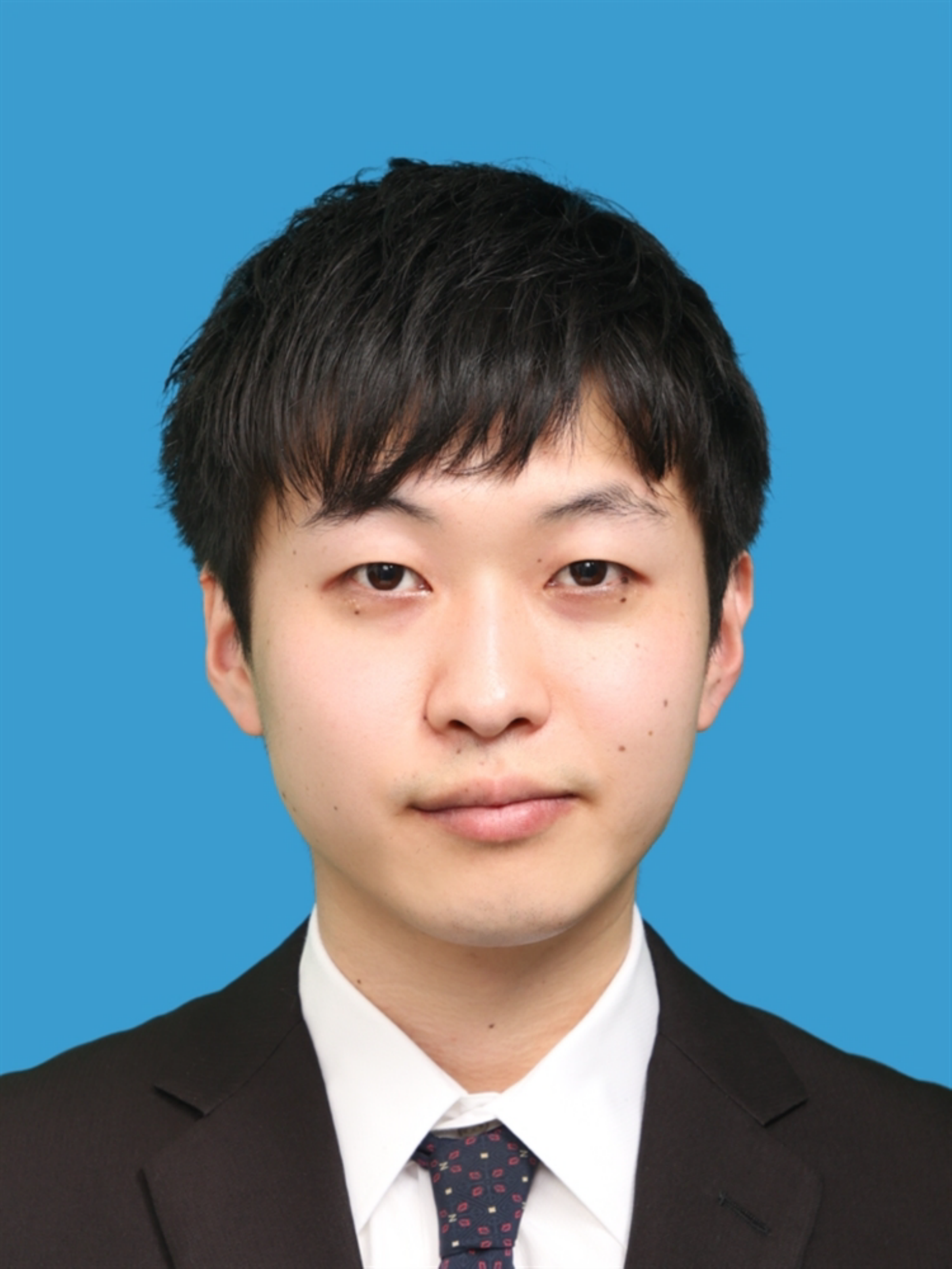}}]{Yuta Takeda}
received the B.S. degree in Informatics and Engineering from The University of Electro-Communications, Tokyo, Japan, in 2022. He is currently an M.S. student at The University of Electro-Communications, Tokyo, Japan.

His research interests include control applications and modeling/control of pneumatic artificial muscles.
\end{IEEEbiography}
\begin{IEEEbiography}[{\includegraphics[width=1in,height=1.25in,clip,keepaspectratio]{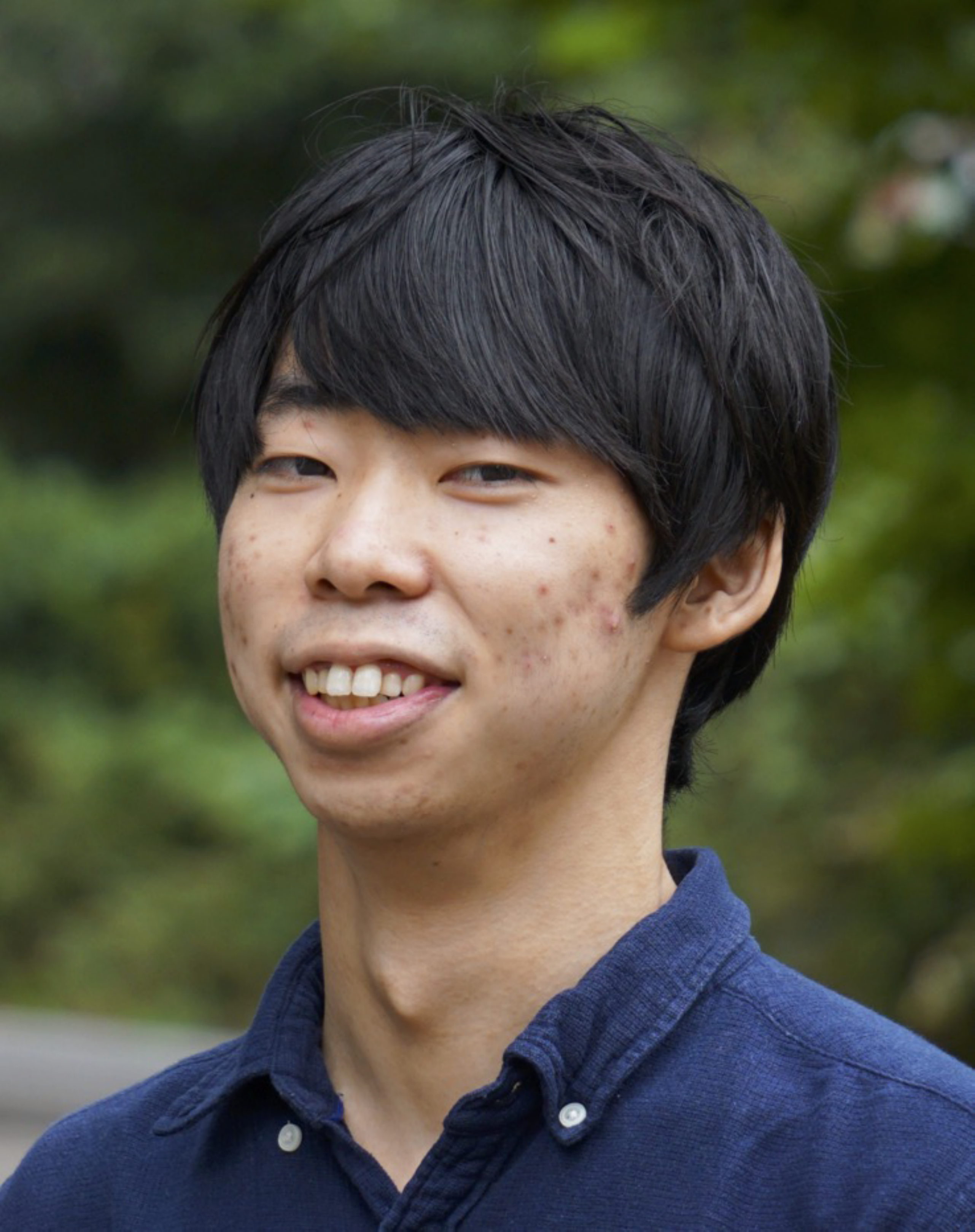}}]{Takaya Shin}
received the B.S. and M.S. degrees in Informatics and Engineering from The University of Electro-Communications, Tokyo, Japan, in 2020 and 2022, respectively. 
He joined Daihen Co., Osaka, Japan.

His research interests include control applications and modeling/control of pneumatic artificial muscles.
\end{IEEEbiography}
\begin{IEEEbiography}[{\includegraphics[width=1in,height=1.25in,clip,keepaspectratio]{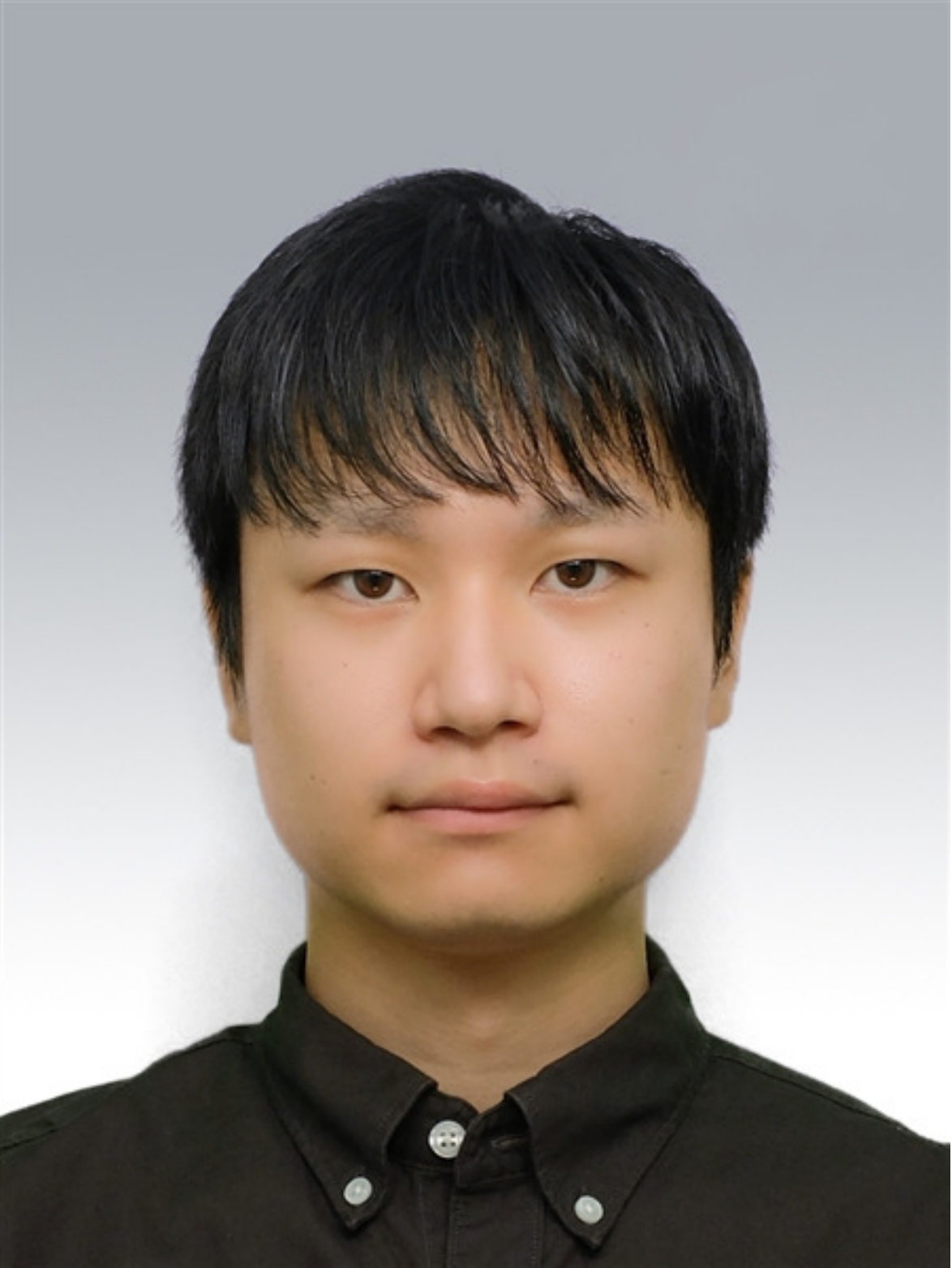}}]
    {Kaoru Teranishi} received the B.S. degree in electromechanical engineering from National Institute of Technology, Ishikawa College, Ishikawa, Japan, in 2019.
    He also obtained the M.S. degree in Mechanical and Intelligent Systems Engineering from The University of Electro-Communications, Tokyo, Japan, in 2021.
    He is currently a Ph.D. student at The University of Electro-Communications.
    From October 2019 to September 2020, he was a visiting scholar of the Georgia Institute of Technology, GA, USA.
    Since April 2021, he has been a Research Fellow of the Japan Society for the Promotion of Science.
    His research interests include control theory and cryptography for cyber-security of control systems.
\end{IEEEbiography}
\begin{IEEEbiography}[{\includegraphics[width=1in,height=1.25in,clip,keepaspectratio]{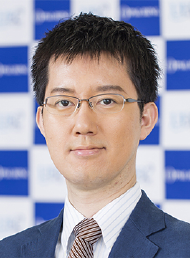}}]{Kiminao Kogiso}
received the B.E., M.E., and Ph.D. degrees in mechanical engineering from Osaka University, Japan, in 1999, 2001, and 2004, respectively. 
He was appointed a postdoctoral fellow in the 21st Century COE Program and as an Assistant Professor in the Graduate School of Information Science, Nara Institute of Science and Technology, Nara, Japan, in April 2004 and July 2005, respectively.
From November 2010 to December 2011, he was a visiting scholar at the Georgia Institute of Technology, GA, USA.
In March 2014, he was promoted to associate professor in the Department of Mechanical and Intelligent Systems Engineering at The University of Electro-Communications, Tokyo, Japan. 
Since April 2023, he has been a professor in the same department.
His research interests include the cybersecurity of control systems, constrained control, control of decision-makers, and their applications.
\end{IEEEbiography}

\vfill

\end{document}